\documentclass[a4paper,11pt]{article}
\usepackage{tablefootnote}
\usepackage[english]{babel}
\usepackage[utf8x]{inputenc}
\usepackage[T1]{fontenc}
\usepackage[flushmargin]{footmisc}
\usepackage{setspace}
\usepackage[comma]{natbib}
\usepackage{array}
\usepackage{float}
\usepackage{amsmath}
\usepackage{amsfonts}
\usepackage{amssymb}
\usepackage{ae}
\usepackage{caption}
\usepackage[a4paper,left=2.5cm,right=2.5cm,top=3cm,bottom=3cm]{geometry}
\usepackage{graphicx}
\usepackage[colorinlistoftodos]{todonotes}
\usepackage[colorlinks=true, allcolors=blue]{hyperref}
\begin{document} \doublespacing \pagestyle{plain}

\def\ci{\perp\!\!\!\perp}
\begin{center}

{\LARGE Causal mediation analysis with double machine learning}

{\large \vspace{0.15cm}}

{ Helmut Farbmacher+, Martin Huber*,  Luk\'{a}\v{s} Laff\'{e}rs++, Henrika Langen*, Martin Spindler**}\medskip

{\small {+Max Planck Society, Munich Center for the Economics of Aging \\ *University of Fribourg, Dept.\ of Economics\\
++Matej Bel University, Dept.\ of Mathematics\\ **University of Hamburg, Faculty of Business Sciences} \bigskip }
\end{center}

\vspace{0.15cm}

\noindent \textbf{Abstract:} {\small This paper combines causal mediation analysis with double machine learning to control for observed confounders in a data-driven way under a selection-on-observables assumption in a high-dimensional setting. We consider the average indirect effect of a binary treatment operating through an intermediate variable (or mediator) on the causal path between the treatment and the outcome, as well as the unmediated direct effect. Estimation is based on efficient score functions, 
which possess a multiple robustness property w.r.t.\ misspecifications of the outcome, mediator, and treatment models. This property is key for selecting these models by double machine learning, 
which is combined with data splitting to prevent overfitting in the estimation of the effects of interest. We demonstrate that the direct and indirect effect estimators are asymptotically normal and root-$n$ consistent under specific regularity conditions and investigate the finite sample properties of the suggested methods in a simulation study when considering lasso as machine learner. We also provide an empirical application to the U.S.\ National Longitudinal Survey of Youth, assessing the indirect effect of health insurance coverage on general health operating via routine checkups as mediator, as well as the direct effect. We find a moderate short-term effect of health insurance coverage on general health which is, however, not mediated by routine checkups.}

{\small \smallskip }



{\small \smallskip {\tiny 
Addresses for correspondence: Helmut Farbmacher, Max Planck Society, Munich Center for the Economics of Aging, Amalienstr. 33, 80799 Munich, Germany; farbmacher@mpisoc.mpg.de.  Martin Huber, University of Fribourg, Department  of  Economics, Bd.\ de P\'{e}rolles 90, 1700 Fribourg, Switzerland; martin.huber@unifr.ch. Luk\'{a}\v{s} Laff\'{e}rs, Matej Bel University, Department of Mathematics, Tajovskeho 40, 97411 Bansk\'{a} Bystrica, Slovakia; lukas.laffers@gmail.com. Henrika Langen, University of Fribourg, Department  of  Economics, Bd.\ de P\'{e}rolles 90, 1700 Fribourg, Switzerland; henrika.langen@unifr.ch. Martin Spindler, University of Hamburg, Faculty of Business Administration, Moorweidenstr.\ 18, 20148 Hamburg, Germany; martin.spindler@uni-hamburg.de. Laff\'{e}rs acknowledges support provided by the Slovak Research and Development Agency under contract no.\ APVV-17-0329 and VEGA-1/0692/20.
}\thispagestyle{empty}\pagebreak  }

{\small \renewcommand{\thefootnote}{\arabic{footnote}} %
\setcounter{footnote}{0}  \pagebreak \setcounter{footnote}{0} \pagebreak %
\setcounter{page}{1} }

\section{Introduction}\label{intro}

Causal  mediation analysis aims at decomposing the causal effect of a treatment on an outcome of interest into an indirect effect operating through a mediator (or intermediate outcome) and a direct effect comprising any causal mechanisms not operating through that mediator. Even if the treatment is random, direct and indirect effects are generally not identified by naively controlling for the mediator without accounting for its likely endogeneity, see \cite{RoGr92}. While much of the earlier literature either neglected endogeneity issues or relied on restrictive linear models, see for instance \cite{Co57}, \cite{JuKe81}, and \cite{BaKe86}, more recent contributions consider more general identification approaches using the potential outcome framework. Some of the numerous examples are \cite{RoGr92}, \cite{Pearl01}, \cite{Robins2003},  \cite{PeSiva06}, \cite{VanderWeele09}, \cite{ImKeYa10}, \cite{Hong10}, \cite{AlbertNelson2011},
\cite{ImYa2011},  \cite{TchetgenTchetgenShpitser2011}, \cite{VansteelandtBekaertLange2012}, and \cite{Huber2012}. Using the denomination of \cite{Pearl01}, the literature distinguishes between natural direct and indirect effects, where mediators are set to their potential values `naturally' occurring under a specific treatment assignment, and the controlled direct effect, where the mediator is set to a `prescribed' value.

The vast majority of identification strategies relies on selection-on-observable-type assumptions implying that the treatment and the mediator are conditionally exogenous when controlling for observed covariates. Empirical examples in economics and policy evaluation include \cite{FlFl09}, \cite{HeckmanPintoSavelyev2013}, \cite{Keeleetal2015}, \cite{ContiHeckmanPinto2016}, \cite{Huber2015}, \cite{HuberLechnerMellace2017},  \cite{BellaniBia2018}, \cite{BijwaardJones2018}, and \cite{HuberLechnerStrittmatter2018}. Such studies typically rely on the (implicit) assumption that the covariates to be controlled for can be unambiguously preselected by the researcher, for instance based on institutional knowledge or theoretical considerations. This assumes away uncertainty related to model selection w.r.t.\ covariates to be included and entails incorrect inference under the common practice of choosing and refining the choice of covariates based on their predictive power.

To improve upon this practice, this paper combines causal mediation analysis based on efficient score functions, see \cite{TchetgenTchetgenShpitser2011}, with double machine learning as outlined in \cite{Chetal2018} for a data-driven control of observed confounders to obtain valid inference under specific regularity conditions. In particular, one important condition is that the number of important confounders (that make the selection-on-observables assumptions to hold approximately) is not too large relative to the sample size. However, the set of these important confounders need not be known a priori and the set of potential confounders can be even larger than the sample size.\footnote{Different from conventional semiparametric methods, the double machine learning framework does not require the set of potential confounders to be restricted by Donsker conditions, but permits the set to be unbounded and to grow with the sample size.} This is particularly useful in high dimensional data with a vast number of covariates that could potentially serve as control variables, which can render researcher-based covariate selection complicated if not infeasible. We demonstrate root-$n$ consistency and asymptotic normality of the proposed effect estimators under specific regularity conditions by verifying that the general framework of \cite{Chetal2018} for well-behaved double machine learning is satisfied in our context.

\cite{TchetgenTchetgenShpitser2011} suggest estimating natural direct and indirect effects based on the efficient score functions of the potential outcomes, which requires plug-in estimates for the conditional mean outcome, mediator density, and treatment probability. Analogous to doubly robust estimation of average treatment effects, see  \cite{Robins+94} and \cite{RoRo95}, the resulting estimators are semiparametrically efficient if all models of the plug-in estimates are correctly specified and remain consistent even if one model is misspecified. Our first contribution is to show that the efficient score function of \cite{TchetgenTchetgenShpitser2011} satisfies the so-called \cite{Neyman1959} orthogonality discussed in \cite{Chetal2018}, which makes the estimation of direct and indirect effects rather insensitive to (local) estimation errors in the plug-in estimates. Second, we show that by an application of Bayes' Theorem, the score function of \cite{TchetgenTchetgenShpitser2011} can be transformed in a way that avoids estimation of the conditional mediator density and show it to be Neyman orthogonal. This appears particularly useful when the mediator is a vector of variables and/or continuous. Third, we establish the score function required for estimating the controlled direct effect along with Neyman orthgonality.

Neyman orthgonality is key for the fruitful application of double machine learning, ensuring robustness in the estimation of the nuisance parameters which is crucial when applying modern machine learning methods. Random sample splitting -- to estimate the parameters of the plug-in models in one part of the data, while predicting the score function and estimating the direct and indirect effects in the other part --  avoids overfitting the plug-in models (e.g.\ by controlling for too many covariates). It increases the variance by only using part of the data for effect estimation. This is avoided by cross-fitting which consists of swapping the roles of the data parts for estimating the plug-in models and the treatment effects to ultimately average over the effects estimates in either part. When combining efficient score-based effect estimation with sample splitting, $n^{-1/2}$-convergence  of treatment effect estimation can be obtained under a substantially slower convergence of $n^{-1/4}$ for the plug-in estimates, see \cite{Chetal2018}. Under specific regularity conditions, this convergence rate can attained by various machine learning algorithms including lasso regression, see \cite{Tibshirani96}.

We investigate the estimators' finite sample behavior based on the score function of \cite{TchetgenTchetgenShpitser2011} and the alternative score suggested in this paper when using post-lasso regression as machine learner for the plug-in estimates. Furthermore, we apply our method to data from the National Longitudinal Survey of Youth 1997 (NLSY97), where a large set of potential control variables is available. We disentangle the short-term effect of health insurance coverage on general health into an indirect effect which operates via the incidence of a routine checkup in the last year and a direct effect covering any other causal mechanisms. While we find a moderate health-improving direct effect, the indirect effect is very close to zero. We therefore do not find evidence that health insurance coverage affects general health through routine checkups in the short run.

We note that basing estimation on efficient score functions is not the only framework satisfying the previously mentioned robustness w.r.t.\ estimation errors in plug-in parameters. This property
is also satisfied by the targeted maximum likelihood estimation (TMLE) framework by \cite{vanderLaanRubin2006}, see the discussion in \cite{Diaz2020}. TMLE relies on iteratively updating (or robustifying) an initial estimate of the parameter of interest based on regression steps that involve models for the plug-in parameters.  \cite{ZhengvanderLaan2012} have developed an estimation approach for natural direct and indirect effects using TMLE, where the plug-in parameters might by estimated by machine learners, e.g.\ the super learner, an ensemble method suggested by \cite{vanderLaanetal2007}. This iterative estimation approach is therefore an alternative to the double machine learning-based approach suggested in this paper, for which we demonstrate $n^{-1/2}$-consistency under specific conditions.

This paper proceeds as follows. Section \ref{effdef} introduces the concepts of direct and indirect effect identification in the potential outcome framework. In Section \ref{ident}, we present the identifying assumptions and discuss identification based on efficient score functions. Section \ref{section:CrossFitting} proposes an estimation procedure based on double machine learning and shows root-$n$ consistency and asymptotic normality under specific conditions. Section \ref{Sim} provides a simulation study. Section \ref{Application} presents an empirical application to data from the NLSY97. Section \ref{conclusion} concludes.

\section{Definition of direct and indirect effects}\label{effdef}

We aim at decomposing the average treatment effect (ATE) of a binary treatment, denoted by $D$, on an outcome of interest, $Y$, into an indirect effect operating through a  discrete mediator, $M$, and a direct effect that comprises any causal mechanisms other than through $M$. We use the potential outcome framework, see for instance  \cite{Rubin74}, to  define the direct and indirect effects of interest, see also \cite{TenHaveetal2007} and \cite{Albert2008} for further examples in the context of mediation. $M(d)$ denotes the potential mediator under treatment value $d$ $\in$ $\{0,1\}$, while $Y(d,m)$ denotes the potential outcome as a function of both the treatment and some value $m$ of the mediator $M$.\footnote{Throughout this paper, capital letters denote random variables and small letters specific values of random variables.} The observed outcome and mediator correspond to the respective potential variables associated with the actual treatment assignment, i.e.\ $Y=D\cdot Y(1,M(1)) + (1-D)\cdot Y(0,M(0))$ and $M=D\cdot M(1) + (1-D)\cdot M(0)$, implying that any other potential outcomes or mediators are a priori (i.e.\ without further statistical assumptions) unknown.

We denote the ATE by $\Delta=E[Y(1,M(1))-Y(0,M(0))]$, which comprises both direct and indirect effects. To decompose the latter, note that the average direct effect, denoted by $\theta (d)$, equals the difference in mean potential outcomes when switching the treatment while keeping the potential mediator fixed, which blocks the causal mechanism via $M$:
\begin{eqnarray}
\theta (d) &=&E[Y(1,M(d))-Y(0,M(d))],\quad d\in\{0,1\}.
\end{eqnarray}
The (average) indirect effect, $\delta (d)$, equals the difference in mean potential outcomes when switching the potential mediator values while keeping the treatment fixed to block the direct effect.
\begin{eqnarray}
\delta (d) &=&E[Y(d,M(1))-Y(d,M(0))],\quad d\in\{0,1\}.
\end{eqnarray}
\cite{RoGr92} and \cite{Robins2003} referred to these parameters as pure/total direct and indirect effects, \cite{FlFl09} as net and mechanism average treatment effects, and \cite{Pearl01} as natural direct and indirect effects, which is the denomination used in the remainder of this paper. %

The ATE is the sum of the natural direct and indirect effects defined upon opposite treatment states $d$, which can be easily seen from adding and subtracting the counterfactual outcomes $E[Y(0,M(1))]$ and $E[Y(1,M(0))]$:
\begin{eqnarray}\label{ate}
\Delta&=&E[Y(1,M(1))-Y(0,M(0))] \notag \\
&=&E[Y(1,M(1))-Y(0,M(1))]+ E[Y(0,M(1))-Y(0,M(0))] = \theta(1) + \delta(0) \notag \\
&=&E[Y(1,M(0))-Y(0,M(0))]+ E[Y(1,M(1))-Y(1,M(0))]=\theta(0) + \delta(1).
\end{eqnarray}
The distinction between $\theta(1)$ and $\theta(0)$ as well as  $\delta(1)$ and $\delta(0)$ hints to the possibility of heterogeneous effects across treatment states $d$ due to interaction effects between $D$ and $M$. For instance, the direct effect of health insurance coverage ($D$) on general health ($Y$) might depend on whether or not a person underwent routine check-ups ($M$). We note that a different approach to dealing with the interaction effects between $D$ and $M$ is a three-way decomposition of the ATE into the pure direct effect ($\theta(0)$), the pure indirect effect ($\delta(0)$) and the mediated interaction effect, see \cite{VanderWeele2013}.

The so-called controlled direct effect, denoted by $\gamma (m)$, is a further parameter that received much attention in the mediation literature. It corresponds to the difference in mean potential outcomes when switching the treatment and fixing the mediator at some value $m$:
\begin{equation}
\gamma (m)=E[Y(1,m)-Y(0,m)],\quad \quad \text{for }m\text{ in the support of }M.
\end{equation}%
In contrast to $\theta (d)$, which is conditional on the potential mediator value `naturally' realized for treatment $d$ which may differ across subjects, $\gamma (m)$ is conditional on enforcing the same mediator state in the entire population. The two parameters are only equivalent in the absence of an interaction between $D$ and $M$. Whether the natural or controlled direct effect is more relevant depends on the feasibility and desirability to intervene on or prescribe the mediator, see \cite{Pearl01} for a discussion of the `descriptive' and `prescriptive' natures of natural and controlled effects. There is no indirect effect parameter matching the controlled direct effect, implying that the difference between the total effect and the controlled direct effect does in general not correspond to the indirect effect, unless there is no interaction between $D$ and $M$, see e.g.\ \cite{Kaufmanetal2004}.

\section{Assumptions and identification}\label{ident}

Our identification strategy is based on the assumption that confounding of the treatment-outcome, treatment-mediator, and mediator-outcome relations can be controlled for by conditioning on observed covariates, denoted by $X$. The latter must not contain variables that are influenced by the treatment, such that $X$ is typically evaluated prior to treatment assignment. Figure \ref{f1} provides a graphical illustration using a directed acyclic graph, with arrows representing causal effects. Each of $D$, $M$, and $Y$  might be causally affected by distinct and statistically independent sets of unobservables not displayed in Figure \ref{f1}, but none of these unobservables may jointly affect two or all three elements $(D,M,Y)$ conditional on $X$.

%
\begin{figure}[!htp]
	\centering \caption{\label{f1}  Causal paths under conditional exogeneity given pre-treatment covariates}\bigskip
\end{figure}

Formally, the first assumption invokes conditional independence of the treatment and potential mediators or outcomes given $X$. This restriction has been referred to as conditional independence, selection on observables, or exogeneity in the treatment evaluation literature, see e.g.\ \cite{Imbens03}. This rules out confounders jointly affecting the treatment on the one hand and the mediator and/or the outcome on the other hand conditional on $X$. In non-experimental data, the plausibility of this assumption critically hinges on the richness of $X$.
\vspace{5pt}\newline
\textbf{Assumption 1 (conditional independence of the treatment):}\newline
$\{Y(d',m), M(d)\}  \bot D | X$ for all $d',d \in \{0,1\}$ and $m$ in the support of $M$,  \vspace{5pt}\newline
where `$\bot$' denotes statistical independence. The second assumption requires the mediator to be conditionally independent of the potential outcomes given the treatment and the covariates.
\vspace{5pt}\newline
\textbf{Assumption 2 (conditional independence of the mediator):}\newline
$ Y(d',m) \bot  M | D=d, X=x  $ for all $d',d \in \{0,1\}$ and $m,x$ in the support of $M,X$.\vspace{5pt}\newline
Assumption 2 rules out confounders jointly affecting the mediator and the outcome conditional on $D$ and $X$.  If $X$ is pre-treatment (as is common to avoid controlling for variables potentially affected by the treatment), this implies the absence of post-treatment confounders of the mediator-outcome relation. Such a restriction needs to be rigorously scrutinized and appears for instance less plausible if the time window between the measurement of the treatment and the mediator is large in a world of time-varying variables.

The third assumption imposes common support on the conditional treatment probability across treatment states.\vspace{5pt}\newline
\textbf{Assumption 3 (common support):}\newline
$\Pr(D=d| M=m, X=x)>0$ for all $d \in \{0,1\}$ and $m,x$ in the support of $M,X$. \vspace{5pt}\newline
The common support assumption, also known as positivity or covariate overlap assumption, restricts the conditional probability to be or not be treated given $M,X$, henceforth referred to as propensity score, to be larger than zero. It implies the weaker condition that $\Pr(D=d|X=x)>0$ such that the treatment must not be deterministic in $X$, otherwise no comparable units in terms of $X$ are available across treatment states. By Bayes' theorem, Assumption 3 also implies that $\Pr(M=m| D=d, X=x)>0$ if $M$ is discrete or that the conditional density of $M$ given $D,X$ is larger than zero if $M$ is continuous. Conditional on $X$, the mediator state must not be deterministic in the treatment, otherwise no comparable units in terms of the treatment are available across mediator states. Assumptions 1 to 3 are standard in the causal mediation literature, see for instance \cite{ImKeYa10}, \cite{TchetgenTchetgenShpitser2011}, \cite{VansteelandtBekaertLange2012}, and \cite{Huber2012}, or also \cite{Pearl01}, \cite{PeSiva06}, and \cite{Hong10}, for closely related restrictions.

\cite{TchetgenTchetgenShpitser2011} discuss identification of the counterfactual $E[Y(d, M(1-d))]$ based on the efficient score function:
\begin{eqnarray}\label{1}
E[Y(d, M(1-d))] &=& E[  \psi_d ],\notag \\
\textrm{ with }\psi_d &=& \frac{I\{D=d\} \cdot f(M|1-d,X)}{p_d(X)\cdot f(M|d,X) }\cdot[Y-\mu(d,M,X)] \notag\\
&& + \frac{I\{D=1-d\}}{1-p_d(X)}\cdot \Big[\mu(d,M,X)  - \int_{m \in \mathcal{M}} \mu(d,m,X)\cdot f(m|1-d,X) \  dm \Big]  \notag\\
&& + \int_{m \in \mathcal{M}}\mu(d,m,X) \cdot f(m|1-d,X) \ dm
\end{eqnarray}
where $ f(M|D,X) $ denotes the conditional density of $ M $ given $ D $ and $ X $ (if $M$ is discrete, this is a conditional probability and integrals need to be replaced by sums), $ p_d(X) = \Pr(D=d|X) $ the probability of treatment $ D=d $ given $ X $, and $ \mu(D,M,X)=E(Y|D,M,X)$ the conditional expectation of outcome $ Y $ given $ D$, $ M $, and $ X $. \eqref{1} satisfies a multiple robustness property in the sense that estimation remains consistent even if one out of the three models for the plug-in parameters $ f(M|D,X) $,  $p_d(X)$, and $\mu(D,M,X)$   is misspecified.

To derive an alternative expression for identification, note that by Bayes' Law,
\begin{eqnarray*}
	\frac{f(M|1-d,X)}{p_d(X)\cdot f(M|d,X)}=\frac{\big(1-p_d(M,X)\big) \cdot  f(M|X)}{  1-p_d(X)}\cdot\frac{ p_{d}(X) }{p_{d}(M,X)  \cdot  f(M|X) \cdot p_d(X)}= \frac{ 1-p_d(M,X) }{ p_{d}(M,X) \cdot \big(1-p_d(X) \big)}
\end{eqnarray*}
where $f(M|X)$ is the conditional distribution of $M$ given $X$ and $ p_d(X,M) = \Pr(D=d|X,M) $. Furthermore,
\begin{eqnarray*}
	&&\int \mu(d,m,X)\cdot f(m|1-d,X) dm=E\Big[ \mu(d,M,X) \Big| D= 1-d, X \Big].
\end{eqnarray*}
Therefore, an alternative  and also multiply robust representation of \eqref{1} is
\begin{eqnarray}\label{2}
	E[Y(d, M(1-d))] &=& E[  \psi_d^* ],\notag \\
	\textrm{ with }
	\psi_d^* &=& \frac{I\{D=d\} \cdot (1-p_d(M,X)) }{ p_{d}(M,X) \cdot  (1-p_d(X))}\cdot[Y-\mu(d,M,X)] \\
	&+& \frac{I\{D=1-d\}}{1-p_d(X)}\cdot \Big[\mu(d,M,X)  - E\Big[ \mu(d,M,X) \Big| D= 1-d, X \Big] \Big]\notag\\  	
	&+& E\Big[ \mu(d,M,X) \Big| D= 1-d, X \Big].\notag
\end{eqnarray}
Similarly as the approaches based on inverse probability weighting (rather than efficient scores) in \cite{Huber2012} and \cite{TchetgenTchetgen2013}, \eqref{2} avoids conditional mediator densities, which appears attractive if $M$ is continuous and/or multidimensional.  On the other hand, it requires the estimation of an additional parameter, namely the nested conditional mean $E[ \mu(d,M,X) | D= 1-d, X ]$, as similarly found in \cite{Milesetal2020}, who suggest a multiply robust score function for assessing path-specific effects. Alternatively to rearranging the score function by \cite{TchetgenTchetgenShpitser2011} as outlined above, ratios of conditional densities as for instance appearing in the first component of  \eqref{1} might be treated as additional nuisance parameter and estimated directly via density-ratio estimation, see e.g. \cite{Sugiyamaetal2010} for density-ratio estimation in high-dimensional settings. Such methods based on directly estimating the density ratio without going through estimating the densities in numerator and denominator separately are shown in several studies to compare favorably with estimating the densities separately, see e.g. \cite{Kanamorietal2012}.

Efficient score-based identification of $E[Y(d, M(d))]$ under $ Y(d,m) \bot  \{D,M\} | X=x$ (see Assumption 1) has been established in the literature on doubly robust ATE estimation, see for instance  \cite{Robins+94} and  \cite{Ha98}:
\begin{eqnarray} \label{meanpot2}
E[Y(d,M(d))]=E[\alpha_d]  \textrm{ with }  \alpha_d=\frac{I\{D=d\}\cdot[Y-\mu(d,X)]}{p_d(X)} + \mu(d,X)
\end{eqnarray}
where $ \mu(D,X)=E(Y|D,M(D),X)=E(Y|D,X)$ is the conditional expectation of outcome $ Y $ given $ D$ and $ X $.

For identifying the controlled direct effect, we now assume that $M$ is discrete (while this need not be the case in the context of natural direct and indirect effects) such that for all $m$ in the support of $M$, it must hold that $\Pr(M=m)>0$. As Assumptions 1 and 2 imply $ Y(d,m) \bot  \{D,M\} | X=x$, doubly robust identification of the potential outcome $E[Y(d,m)]$, which is required for the controlled direct effect, follows from replacing $I\{D=d\}$ and $p_d(X)$ in \eqref{meanpot2} by $I\{D=d, M=m\}= I\{M=m\}\cdot I\{D=d\}$ and $\Pr(D=d,M=m|X)=f(m|d,X)\cdot p_d(X)$:
\begin{eqnarray}\label{conddm}
E[Y(d,m)]=E[ \psi_{dm}]\textrm{ with }\psi_{dm}=\frac{I\{D=d\}\cdot I\{M=m\}\cdot[Y-\mu(d,m,X)]}{ f(m|d,X)\cdot p_d(X)} + \mu(d,m,X).
\end{eqnarray}

\section{Estimation of the counterfactual with K-fold Cross-Fitting}
\label{section:CrossFitting}
We subsequently propose an estimation strategy for the counterfactual  $E[Y(d, M(1-d))]$ with $d \in \{0,1\}$ based on the efficient score function by \cite{TchetgenTchetgenShpitser2011} provided in \eqref{1}  and show its root-$n$ consistency under specific regularity conditions.  To this end, let $\mathcal{W} = \{W_i|1\leq i \leq N\}$ with $W_i = (Y_i, M_i, D_i, X_i)$ for $i=1,\ldots, n$ denote the set of observations in an i.i.d.\ sample of size $n$. $\eta$ denotes the plug-in (or nuisance) parameters, i.e.\ the conditional mean outcome, mediator density and treatment probability. Their respective estimates are referred to by $\hat{\eta} =  \{\hat\mu(D,M,X),\hat{f}(M|D,X), \hat{p_d}(X) \}$ and the true nuisance parameters by $\eta_0 =  \{\mu_0(D,M,X), f_0(M|D,X), p_{d0}(X) \}$. Finally, $\Psi_{d0}=E[Y(d, M(1-d))]$ denotes the true counterfactual.

We suggest estimating $\Psi_{d0}$ using the following algorithm that combines orthogonal score estimation with sample splitting and is root-$n$ consistent under conditions outlined further below. \vspace{5pt}\newline
\textbf{Algorithm 1: Estimation of $E[Y(d, M(1-d))]$ based on equation \eqref{1}}
\begin{enumerate}
	\item Split $\mathcal{W}$ in $ K $ subsamples. For each subsample $ k $, let $n_k$ denote its size, $\mathcal{W}_k$ the set of observations in the sample and $\mathcal{W}_k^{C}$ the complement set of all observations not in $k$.
	\item For each $k$, use  $\mathcal{W}_k^{C}$ to estimate the model parameters of $p_d(X)$, $f(M|D,X)$, and $ \mu(D,M,X) $ in order to predict these models in $\mathcal{W}_k$, where the predictions are denoted by $\hat{p_d}^k(X)$, $\hat{f}^k(M|D,X)$, and $\hat\mu^k(D,M,X)$.
	\item For each $ k $, obtain an estimate of the efficient score function (see $\psi_d$ in \eqref{1}) for each observation $i$ in $\mathcal{W}_k$, denoted by $\hat \psi_{d,i} ^k$ :
	\begin{eqnarray}
	\hat \psi_{d,i}  ^k &=& \frac{I\{D_i=d\} \cdot \hat{f}^k(M_i|1-d,X_i)}{\hat{p}_d^k(X_i)\cdot \hat{f}^k(M_i|d,X_i) }\cdot[Y_i-\hat{\mu}^k(d,M_i,X_i)] \notag\\
	&& + \frac{I\{D_i=1-d\}}{1-\hat{p}_d^k(X_i)}\cdot \Big[\hat{\mu}^k(d,M_i,X_i) - \int_{m \in \mathcal{M}} \hat{\mu}^k(d,m,X_i)\cdot \hat{f}^k(m|1-d,X_i) dm \Big]  \notag\\
	&& + \int_{m \in \mathcal{M}} \hat{\mu}^k(d,m,X_i) \cdot \hat{f}^k(m|1-d,X_i) dm.
	\end{eqnarray}
	\item Average the estimated scores $\hat \psi_{d,i}^k$ over all observations across all $ K $ subsamples to obtain an estimate of  $\Psi_{d0}=E[Y(d, M(1-d))]$ in the total sample, denoted by $\hat \Psi_d=1/n \sum_{k=1}^{K}  \sum_{i=1}^{n_k} \hat \psi_{d,i}^k$.\vspace{5pt}\newline
\end{enumerate}

Algorithm 1 can be adapted to estimate the counterfactuals required for the controlled direct effect, see \eqref{conddm}. To this end, denote by $\Psi_{dm0}=E[Y(d, m)]$ the true counterfactual of interest, which is
estimated by replacing $\psi_{d}$ and $\Psi_d$ by $\psi_{dm}$ and  $\Psi_{dm0}$, respectively, everywhere in Algorithm 1.

In order to achieve root-$n$ consistency for counterfactual estimation, we make specific assumptions about the prediction qualities of the machine learners for our plug-in estimates of the nuisance parameters. Closely following \cite{Chetal2018}, we to this end introduce some further notation. Let $(\delta_n)_{n=1}^{\infty}$ and $(\Delta_n)_{n=1}^{\infty}$ denote sequences of positive constants with $\lim_{n\rightarrow \infty} \delta_n = 0 $ and $\lim_{n\rightarrow \infty} \Delta_n = 0.$ Also, let $c, \epsilon, C, \underline{f},\overline{f}$ and $q$ be positive constants such that $q>2,$ and let $K \geq 2$ be a fixed integer. Furthermore, for any random vector $Z = (Z_1,...,Z_l)$, let $\left\| Z \right\|_{q} = \max_{1\leq j \leq l}\left\| Z_l \right\|_{q},$ where $\left \| Z_l \right\|_{q}  =  \left( E\left[ \left| Z_l \right|^q \right] \right)^{\frac{1}{q}}$. For the sake of easing notation, we assume that $n/K$ is an integer. For brevity, we omit the dependence of probability $\Pr_P(\cdot),$ expectation $E_P(\cdot),$ and norm $\left\| \cdot  \right\|_{P,q}$ on the probability measure $P$. \vspace{5pt}\newline
\textbf{Assumption 4 (regularity conditions and quality of plug-in parameter estimates):}  \newline
For all probability laws $P \in \mathcal{P}$, where $\mathcal{P}$ is the set of all possible probability laws, the following conditions hold for the random vector $(Y,D,M,X)$ for $d \in \{0,1\}$:
\begin{enumerate}
\item[(a)] $  \left\| Y \right\|_{q} \leq C$ and $\left\| E[Y^2|d,M,X] \right\|_{\infty} \leq C^2,$
\item[(b)]

$\Pr(\epsilon \leq p_{d0} (X) \leq 1-\epsilon) = 1,$

\item[(c)]
$\Pr(\underline{f} \leq f(M|D,X) \leq \overline{f})=1,$

\item[(d)]
$\left\| Y-\mu_0(d,M,X)  \right\|_{2} = E_{ } \Big[\left(Y-\mu_0(d,M,X)) \right)^2 \Big]^{\frac{1}{2}} \geq c$

\item[(e)] Given a random subset $I$ of $[n]$ of size $n_k =n/K,$ the nuisance parameter estimator $\hat \eta_0 = \hat \eta_0( \mathcal{W}_k^{C} )$ satisfies the following conditions. With $P$-probability no less than $1-\Delta_n:$
\begin{eqnarray}
\left\|  \hat \eta_0 - \eta_0 \right\|_{q} &\leq& C, \notag \\
\left\|  \hat \eta_0 - \eta_0 \right\|_{2} &\leq& \delta_n, \notag \\
\left\| \hat  p_{d0}(X)-1/2\right\|_{\infty}  &\leq& 1/2-\epsilon, \notag\\
\left\|   \hat f_0(M|D,X)-(\underline{f} + \overline{f})/2 \right\|_{\infty}  &\leq& (\overline{f} - \underline{f})/2, \notag \\
\left\|  \hat \mu_0(D,M,X)-\mu_0(D,M,X)\right\|_{2} \times \left\|  \hat p_{d0}(X)-p_{d0}(X)\right\|_{2}  &\leq & \delta^{}_n n^{-1/2}, \notag \\
\left\|  \hat \mu_0(D,M,X)-\mu_0(D,M,X)\right\|_{2} \times \left\|  \hat f_0(M|1-D,X)-f_{0}(M|1-D,X)\right\|_{2}  &\leq & \delta^{}_n n^{-1/2}. \notag
\end{eqnarray}
\end{enumerate}

For demonstrating root-$n$ consistency of the proposed estimation strategy for the counterfactual, we heavily draw from \cite{Chetal2018}. We show that our estimation strategy satisfies the requirements for their double machine learning framework  by first verifying the satisfaction of a specific moment condition as well as linearity and Neyman orthogonality of the score (see Appendix \ref{Neyman}). Then, as e.g.\ $\psi_d(W, \eta, \Psi_{d0})$ is smooth in $ (\eta,\Psi_{d0}) $, the  plug-in estimators must converge with rate $ n^{-1/4} $ in order to achieve $ n^{-1/2} $-convergence for the estimation of $ \hat \Psi_d $. This convergence rate of $ n^{-1/4} $ is achievable for many commonly used machine learners such as lasso, random forest, boosting and neural nets. The rates for $L_2$-boosting were, for instance, derived in \cite{LuoSpindler2016}.
\vspace{5pt}\newline

\textbf{Theorem 1}\\
Under  Assumptions 1-4, it holds for estimating $E[Y(d, M(1-d))]$, $E[Y(d, m)]$ based on Algorithm 1:  \\
$\sqrt{n} \Big(\hat \Psi_d - \Psi_{d0} \Big) \rightarrow N(0,\sigma^2_{\psi_{d}})$,  where $\sigma^2_{\psi_d}= E[(\psi_{d}-\Psi_{d0})^2]$. \\	
$\sqrt{n} \Big(\hat \Psi_{dm} - \Psi_{dm0} \Big) \rightarrow N(0,\sigma^2_{\psi_{dm}})$, where $\sigma^2_{\psi_d}= E[(\psi_d-\Psi_{dm0})^2]$.\\
The proof is provided in Appendix \ref{proof1}.
\vspace{5pt}\newline

Analogous results follow for the estimation of $\Lambda=E[Y(d, M(d))]$ when replacing $\hat \psi_d$ in the algorithm above by an estimate of score function $\alpha_d$ from \eqref{meanpot2},
\begin{eqnarray}
&\hat \alpha_d=\frac{I\{D=d\} \cdot (Y_i-\hat{\mu}^k(d,X_i))}{\hat{p_d}^k(X_i)}+ \hat{\mu}^k(d,X_i) ,
\end{eqnarray}
where $\hat{\mu}^k(d,x)$ is an estimate of $\mu(d,x)$. This approach has been discussed in literature on ATE estimation based on double machine learning, see for instance \cite{Bellonietal2017} and \cite{Chetal2018}. Denoting by $\hat \Lambda$ the estimate of $\Lambda$, it follows under Assumptions 1-4 that $\sqrt{n} \Big(\hat \Lambda_d - \Lambda_d \Big) \rightarrow N(0,\sigma^2_{\alpha_d})$,
where $\sigma^2_{\alpha_d}= E[(\alpha_d-\Lambda_d)^2]$. Therefore, root-$n$-consistent estimates of the total as well as the direct and indirect effects are obtained as difference of the estimated potential outcomes, which we denote by $\hat \Delta$, $\hat \theta(d)$, and $\hat \delta(d)$. That is, $\hat \Delta=\hat \Lambda_1-\hat \Lambda_0$, $\hat \theta(1)=\hat \Lambda_1-\hat \Psi_0$, $\hat \theta(0)=\hat \Psi_1 - \hat \Lambda_0$, $\hat \delta(1)=\hat \Lambda_1-\hat \Psi_1$, and $\hat \delta(0)=\hat \Psi_0 - \hat \Lambda_0$.

Naturally, the asymptotic variance of any effect is obtained based on the variance of the difference in the score functions of the potential outcomes required for the respective effect. For instance, the asymptotic variance of $\hat \theta(1)$ is given by
$Var(\hat \theta(1))=Var(\alpha_1-\psi_0)/n=(\sigma^2_{\alpha_1}+\sigma^2_{\psi_0}-2Cov(\alpha_1,\psi_0))/n$.

\cite{Chetal2018} show that under Assumptions 1-4, $\hat \sigma^2_{\psi_d}$ can be estimated as:
\begin{eqnarray}
& \hat \sigma^2_{\psi_d} = 1/K \sum_{k = 1}^K  \Big[1/n_k  \sum_{i = 1}^{n_k} \psi_d(W_i, \hat \eta_0^k, \hat \Psi_{d})^2\Big]
\end{eqnarray}
The asymptotic variance of $\alpha_d$ can be estimated accordingly, with $\psi_d$ and $\hat \Psi_{d0}$ substituted by $\alpha_d$ and $\hat \Lambda_{d0}$.

We subsequently discuss estimation based on the score function $\psi_d^*$ in expression \eqref{2}. We note that in this case, one needs to estimate a nested nuisance parameter $E\Big[ \mu(d,M,X) \Big| D= 1-d, X \Big]$. To avoid overfitting, the models for  $\mu(d,M,X)$ and $E\Big[ \mu(d,M,X) \Big| D= 1-d, X \Big]$ are estimated in different subsamples. The plug-in estimates for the conditional mean outcome, mediator density and treatment probability are referred to by $\hat{\eta}^* =  \{ \hat \mu(D,M,X), \hat \omega(D,M,X), \hat p(M,X), \hat p_{d}(X) \}$ and the true nuisance parameters by $\eta_0^* =  \{\mu_0(D,M,X), \omega(D,M,X), p_{d0}(M,X), p_{d0}(X) \}$.
\vspace{5pt}\newline
\textbf{Algorithm 2: Estimation of $E[Y(d, M(1-d))]$ based on equation \eqref{2}}
\begin{enumerate}
	\item Split $\mathcal{W}$ in $ K $ subsamples. For each subsample $ k $, let $n_k$ denote its size, $\mathcal{W}_k$ the set of observations in the sample and $\mathcal{W}_k^{C}$ the complement set of all observations not in $k$.
	\item For each $k$, use  $\mathcal{W}_k^{C}$ to estimate the model parameters of $p_d(X)$ and $p_d(M,X)$.  Split  $\mathcal{W}_k^{C}$ into 2 nonoverlapping subsamples and estimate the model parameters of the conditional mean $\mu(d,M,X)$ and the nested conditional mean $E\Big[ \mu(d,M,X) \Big| D= 1-d, X \Big]$   in the distinct subsamples.  Predict the nuisance parameters in  $\mathcal{W}_k$, where the predictions are denoted by $\hat{p_d}^k(X)$, $\hat{p}_d^k(M,X)$, $\hat\mu^k(D,M,X)$ and $\hat{E}\Big[ \mu(d,M,X)\ \Big| D= 1-d, X \Big]^k$.
	\item For each $ k $, obtain an estimate of the efficient score function (see $\psi_d^*$ in \eqref{2}) for each observation $i$ in $\mathcal{W}_k$, denoted by $\hat \psi_{d,i} ^k$ :
	\begin{eqnarray}
	\hat \psi_{d,i}^{*k} &=&  \frac{I\{D_i=d\} \big( 1-\hat{p}_d^k(M_i,X_i) \big) }{ \hat{p}_d^k(M_i,X_i) \, \big( 1-\hat{p}_d^k(X_i) \big) }\cdot[Y-\hat\mu^k(d,M_i,X_i)]\notag \\
	&& +\frac{I\{D_i=1-d\}}{1-\hat{p}_d^k(X_i)}\cdot \Big[\hat\mu^k(d,M_i,X_i)  -  \hat{E}\Big[ \hat\mu^k(d,M_i,X_i) \Big| D_i=1-d, X_i \Big] \Big]\notag\\
	&&  + \hat{E}\Big[ \hat\mu^k(d,M_i,X_i) \Big| D_i=1-d, X_i \Big].
	\end{eqnarray}
	\item Average the estimated scores $\hat \psi_{d,i}^{*k}$ over all observations across all $ K $ subsamples to obtain an estimate of  $\Psi_{d0}=E[Y(d, M(1-d))]$ in the total sample, denoted by $\hat \Psi_d^*=1/n \sum_{k=1}^{K}  \sum_{i=1}^{n_k} \hat \psi_{d,i}^{*k}$.\vspace{5pt}\newline
\end{enumerate}
Also this approach can be shown to be root-$n$-consistent under specific regularity conditions outlined below.\vspace{5pt}\newline
\textbf{Assumption 5 (regularity conditions and quality of plug-in parameter estimates):}  \newline
For all probability laws $P \in \mathcal{P}$
 the following conditions hold for the random vector $(Y,D,M,X)$ for all $d \in \{0,1\}$:
\begin{enumerate}
\item[(a)] $  \left\| Y \right\|_{q} \leq C$ and $\left\| E[Y^2|d,M,X] \right\|_{\infty} \leq C^2,$

\item[(b)]

$\Pr(\epsilon \leq p_{d0} (X) \leq 1-\epsilon) = 1,$

\item[(c)]
$\Pr(\epsilon \leq p_{d0} (M,X) \leq 1-\epsilon) = 1,$

\item[(d)]
$\left\| Y-\mu_0(d,M,X)  \right\|_{2} = E_{ } \Big[\left(Y-\mu_0(d,M,X)) \right)^2 \Big]^{\frac{1}{2}} \geq c$

\item[(e)] Given a random subset $I$ of $[n]$ of size $n_k =n/K,$ the nuisance parameter estimator $\hat \eta^*_0 = \hat \eta^*_0( \mathcal{W}_k^{C} )$ satisfies the following conditions. With $P$-probability no less than $1-\Delta_n:$
\begin{eqnarray}
\left\|  \hat \eta^*_0 - \eta^*_0 \right\|_{q} &\leq& C, \notag \\
\left\|  \hat \eta^*_0 - \eta^*_0 \right\|_{2} &\leq& \delta_n, \notag \\
\left\| \hat  p_{d0}(X)-1/2\right\|_{\infty}  &\leq& 1/2-\epsilon, \notag\\
\left\| \hat  p_{d0}(M,X)-1/2\right\|_{\infty}  &\leq& 1/2-\epsilon, \notag\\
\left\| \hat \mu_0(D,M,X)-\mu_0(D,M,X)\right\|_{2} \times \left\|  \hat p_{d0}(X)-p_{d0}(X)\right\|_{2}  &\leq & \delta^{}_n n^{-1/2}, \notag\\
\left\|  \hat \mu_0(D,M,X)-\mu_0(D,M,X)\right\|_{2} \times \left\|  \hat p_{d0}(M,X)-p_{d0}(M,X)\right\|_{2}  &\leq & \delta^{}_n n^{-1/2}, \notag\\
\left\|  \hat \omega_0(D,M,X)-\omega_0(D,M,X)\right\|_{2} \times \left\|  \hat p_{d0}(X)-p_{d0}(X)\right\|_{2}  &\leq & \delta^{}_n n^{-1/2}. \notag
\end{eqnarray}
\end{enumerate}
\vspace{5pt}
\textbf{Theorem 2}\\
Under  Assumptions 1-3 and 5, it holds for estimating $E[Y(d, M(1-d))]$ based on Algorithm 2:  \\
$\sqrt{n} \Big(\hat \Psi_d^* - \Psi_{d0}^* \Big) \rightarrow N(0,\sigma^2_{\psi_{d}^*})$, where $\sigma^2_{\psi_d^*}= E[(\psi_{d}^*-\Psi_{d0}^*)^2]$.\\
The proof is provided in Appendix \ref{altscoreproof}.

\section{Simulation study}\label{Sim}
This section provides a simulation study to investigate the finite
sample behaviour of the proposed methods based on the following data
generating process:
\begin{eqnarray*}
	Y&=&0.5  D + 0.5 M + 0.5 D  M + X'\beta + U ,\\
	M&=&I\{0.5 D + X'\beta + V>0\},\quad D=I\{X'\beta+ W>0\},\\
	X &\sim& N(0,\Sigma),\quad U, V, W \sim N(0,1)\textrm{ independently of each other and $X$}.
\end{eqnarray*}

Outcome $Y$ is a function of the observed variables $D,M,X$, including an interaction between the mediator and the treatment, and an
unobserved term $U$.  The binary mediator $M$ is a function of $D,X$ and the unobservable $V$, while the binary treatment $D$ is  determined by $X$ and the unobservable $W$. $X$ is a vector of covariates of dimension $p$, which is drawn from a multivariate normal distribution with zero mean and covariance matrix $\Sigma$. The latter is defined based on setting the covariance of the $i$th and $j$th covariate in $X$ to $\Sigma_{ij}=0.5^{|i-j|}$.\footnote{The results presented below are hardly affected when setting  $\Sigma$ to the identity matrix (zero correlation across $X$).} Coefficients $\beta$  gauge the impact of $X$ on $Y$, $M$, and $D$, respectively, and thus, the strength of confounding. $ U, V, W$ are random and standard normally distributed scalar unobservables.
We consider two sample sizes of $n=1000,4000$ and run $1000$ simulations per data generating process.

We investigate the performance of effect estimation based on (i) Theorem 1 using the identification result in expression \eqref{1} derived by \cite{TchetgenTchetgenShpitser2011} as well as (ii) Theorem 2 using the modified score function  in expression \eqref{2} which avoids conditional mediator densities. 
The nuisance parameters are estimated by post-lasso regression based on the `hdm' package by \citet{Rhdm} for the statistical software `R' with its default options, using logit specifications for $p_d(X)$, $p_d(M,X)$, and $f(M|D,X)$ and linear specifications for $\mu(D,M,X)$ and $E[ \mu(d,M,X)| D=1-d, X ]$. The estimation of direct and indirect effects is based on 3-fold cross-fitting. For all methods investigated, we drop observations whose (products of) estimated conditional probabilities in the denominator of any potential outcome expression are close to zero, namely smaller than a trimming threshold of $0.05$ (or 5\%). Our estimation procedure is available in the `causalweight' package for `R' by \citet{BodoryHuber2018}.

In our first simulation design, we set $p=200$ and the $i$th element in the coefficient vector $\beta$ to $0.3/i^2$ for $i=1,...,p$, meaning a quadratic decay of covariate importance in terms of confounding.  This specification implies that the $R^2$ of $X$ when predicting $Y$ amounts to 0.22 in large samples, while the \cite{Nagelkerke1991} pseudo-$R^2$ of $X$ when predicting $D$ and $M$ by probit models amounts to  0.10 and 0.13, respectively. The left panel of Table \ref{table:sim1} reports the results for either sample size. For $n=1000$, double machine learning based on Theorem 2 on average exhibits a slightly lower absolute bias (`abias') and standard deviation (`sd') than estimation based on Theorem 1. The behavior of both approaches improves when increasing sample size to $n=4000$, as the absolute bias is very close to zero for any effect estimate and standard deviation is roughly cut by half. Under the larger sample size, differences in terms of root mean squared error (`rmse') between estimation based on Theorems 1 and 2 are very close to zero. By and large, the results suggest that the estimators converge to the true effects at root-$n$ rate.

In our second simulation, confounding is increased by setting $\beta$ to $0.5/i^2$ for $i=1,...,p$. This specification implies that the $R^2$ of $X$ when predicting $Y$ amounts to 0.42, while the \cite{Nagelkerke1991} pseudo-$R^2$ of $X$ when predicting $D$ and $M$ amounts to 0.23 and 0.28, respectively. The results are displayed in the right panel of Table \ref{table:sim1}.  Again, estimation based on Theorem 2 slightly dominates in terms of having a smaller absolute bias and standard deviation, in particular for $n=1000$.  However, in other settings, the two methods might compare differently in terms of finite sample performance. Both methods based on Theorems 1 and 2, respectively, appear to converge to the true effects at root-$n$ rate, and differences in terms of root mean squared errors are minor for $n=4000$.

\begin{table}[H]
	\caption{Simulation results for effect estimates ($p=200$)}\label{table:sim1}
	{\footnotesize
		\begin{center}
			\begin{tabular}{r|ccc|ccc|c|ccc|ccc|c}
				\hline\hline
				& \multicolumn{7}{c|}{ Coefficients given by $0.3/i^2$ for $i=1,...,p$} & \multicolumn{7}{c}{Coefficients given by $0.5/i^2$ for $i=1,...,p$ }  \\
				& abias & sd & rmse & abias & sd & rmse & true & abias & sd & rmse & abias & sd & rmse & true \\
				& \multicolumn{3}{c|}{$n$=1000} & \multicolumn{3}{c|}{$n$=4000} &  & \multicolumn{3}{c|}{$n$=1000} & \multicolumn{3}{c|}{$n$=4000} & \\
				\hline
				& \multicolumn{14}{c}{Double machine learning based on Theorem 1}  \\
				\hline
				$\hat{\Delta}$  & 0.01 & 0.08 & 0.08 & 0.00 & 0.04 & 0.04 & 1.02 & 0.02 & 0.09 & 0.09 & 0.02 & 0.04 & 0.05 & 1.00 \\
				$\hat{\theta}(1)$ & 0.00 & 0.09 & 0.09 & 0.00 & 0.04 & 0.04 & 0.84 & 0.01 & 0.09 & 0.09 & 0.01 & 0.04 & 0.04 & 0.83 \\
				$\hat{\theta}(0)$  & 0.01 & 0.08 & 0.08 & 0.00 & 0.04 & 0.04 & 0.75 & 0.02 & 0.08 & 0.09 & 0.01 & 0.04 & 0.04 & 0.75 \\
				$\hat{\delta}(1)$ & 0.00 & 0.06 & 0.06 & 0.00 & 0.03 & 0.03 & 0.27 & 0.00 & 0.06 & 0.06 & 0.00 & 0.03 & 0.03 & 0.25 \\
				$\hat{\delta}(0)$ & 0.01 & 0.06 & 0.06 & 0.00 & 0.02 & 0.02 & 0.18 & 0.01 & 0.06 & 0.06 & 0.00 & 0.02 & 0.02 & 0.17 \\
				trimmed &   \multicolumn{3}{c|}{ 17.24  } &  \multicolumn{3}{c|}{ 19.19 } & &   \multicolumn{3}{c|}{ 80.25  } &  \multicolumn{3}{c|}{ 237.50  } &  \\
				\hline
					& \multicolumn{14}{c}{Double machine learning based on Theorem 2}  \\ 
					\hline	
					$\hat{\Delta}$ & 0.00 & 0.08 & 0.08 & 0.00 & 0.04 & 0.04 & 1.02 & 0.01 & 0.09 & 0.09 & 0.01 & 0.04 & 0.04 & 1.00 \\
					$\hat{\theta}(1)$ & 0.00 & 0.08 & 0.08 & 0.00 & 0.04 & 0.04 & 0.84 & 0.00 & 0.08 & 0.08 & 0.00 & 0.04 & 0.04 & 0.83 \\
						$\hat{\theta}(0)$ & 0.00 & 0.08 & 0.08 & 0.00 & 0.04 & 0.04 & 0.75 & 0.00 & 0.08 & 0.08 & 0.00 & 0.04 & 0.04 & 0.75 \\
					$\hat{\delta}(1)$ & 0.00 & 0.06 & 0.06 & 0.00 & 0.03 & 0.03 & 0.27 & 0.00 & 0.06 & 0.06 & 0.00 & 0.03 & 0.03 & 0.25 \\
					$\hat{\delta}(0)$ & 0.00 & 0.04 & 0.04 & 0.00 & 0.02 & 0.02 & 0.18 & 0.00 & 0.05 & 0.05 & 0.00 & 0.02 & 0.02 & 0.17 \\
			trimmed &   \multicolumn{3}{c|}{  1.20 } &  \multicolumn{3}{c|}{ 0.11   } & &   \multicolumn{3}{c|}{ 16.76  } &  \multicolumn{3}{c|}{ 25.45 } &  \\
\hline
\end{tabular}
\end{center}
\par
Note: `abias', `sd', and `rmse' denote the absolute bias, standard deviation and root mean squared error of the respective effect estimate. `true' provides the true effect. `trimmed' is the average number of trimmed observations per simulation. The propensity score-based trimming threshold is set to $0.05$.
}
\end{table}

Appendix \ref{simse} reports the simulation results (namely the absolute bias, standard deviation, and root mean squared error) for the standard errors obtained by an asymptotic approximation based on the estimated variance of the score functions. The results suggest that the asymptotic standard errors decently estimate the actual standard deviation of the point estimators.

\section{Application}\label{Application}

In this section, we apply our method to data from the National Longitudinal Survey of Youth 1997 (NLSY97), a survey following a U.S. nationally representative sample of  8,984 individuals born in the years 1980-84. Since 1997, the participants have been interviewed on a wide range of demographic, socioeconomic, and health-related topics in a one- to two-year circle. We investigate the causal effect of health insurance coverage ($D$) on general health ($Y$) and decompose it into an indirect pathway via the incidence of a regular medical checkup ($M$) and a direct effect entailing any other causal mechanisms. Whether or not an individual undergoes routine checkups appears to be an interesting mediator, as it is likely to be affected by health insurance coverage and may itself have an impact on the individual's health, because checkups can help identifying medical conditions before they get serious to prevent them from affecting a person's general health state. 

The effect of health insurance coverage on self-reported health has been investigated in different countries with no compulsory medical insurance and no publicly provided universal health coverage, see for example \cite{simon2017impact}, \cite{sommers2017three}, \cite{baicker2013oregon}, \cite{yoruk2016health} and \cite{cardella2014effect} for the U.S.\ and \cite{king2009public} for Mexico). Most of these studies find a significant positive effect of insurance coverage on self-reported health. The impact of insurance coverage on the utilization of preventive care measures, particularly routine checkups like cancer, diabetes and cardiovascular screenings, is also extensively covered in public health literature. Most studies find that health insurance coverage increases the odds of attending routine checkups. While some contributions include selected demographic, socioeconomic and health-related control variables to account for the endogeneity of health insurance status (see e.g. \cite{faulkner1997effect}, \cite{press2014insurance}, \cite{burstin1998effect}, \cite{fowler2007risk}),  others exploit natural experiments: \cite{simon2017impact} estimate a difference-in-differences model comparing states which did and did not expand Medicaid to low-income adults in 2005, while \cite{baicker2013oregon} exploit that the state of Oregon expanded Medicaid based on lottery drawings from a waiting list. The results of both studies suggest that the Medicaid expansions increased use of certain forms of preventive care. In a study on Mexican adults, \cite{pagan2007health} use self-employment  and  commission pay as instruments for insurance coverage and also find a more frequent use of some types of preventive care by individuals with health insurance coverage.

While the bulk of studies investigating checkups focus on one particular type of screening (rather than general health checkups), see \cite{maciosek2010greater} for a literature review, several experimental contributions also assess general health checkups. For instance, \cite{rasmussen2007preventive} conduct an experiment with individuals aged 30 to 49 in Denmark by randomly offering a set of health screenings, including advice on healthy living and find a significant positive effect on life expectation. In a study on Japan’s elderly population, \cite{nakanishi1996preventive} find a significantly negative correlation between the rate of attendance at health check-ups and hospital admission rates. Despite the effects of health insurance coverage and routine checkups being extensively covered in the public health literature, the indirect effect of insurance on general health operating via routine checkups as mediator has to the best of our knowledge not yet been investigated. 
A further distinction to most previous studies is that we consider comparably young individuals with an average age below 30. For this population, the relative importance of different health screenings might differ from that for other age groups. We also point out that our application focuses on short-term health effects.


We consider a binary indicator for health insurance coverage, equal to one if an individual reports to have any kind of health insurance when interviewed in 2006 and zero otherwise. The outcome, self-reported general health, is obtained from 2008 interview and measured with an ordinal variable, taking on the values `excellent', `very good', `good', `fair' and `poor'. In the 2007 interview, participants were asked whether they have gone for routine checkups since the 2006 interview. This information serves as binary mediator, measured post-treatment but pre-outcome.

To ensure that the control variables ($X$) are not influenced by the treatment, they come from the pre-treatment 2005 and earlier interview rounds. They cover demographic characteristics, family background and quality of the home environment during youth, education and training, labor market status, income and work experience, marital status and fertility, household characteristics, received monetary transfers, attitudes and expectations, state of physical and mental health as well as health-related behavior regarding e.g. nutrition and physical activity. For some variables, we only consider measurements from 2005 or from the initial interview round covering demographics and family related topics. For other variables we include measurements from both the indiviuals' youth and 2005 in order to capture their social, emotional and physical development. Treatment and mediator state in the pre-treatment period (2005) are also considered as potential control variables. Item non-response in control variables is dealt with by including missing dummies for each control variable and setting the respective missing values to zero. In total, we end up with a set of 770 control variables, 601 of which are dummy variables (incl. 252 dummies for missing values).

After excluding 1923 observations with either mediator or treatment status missing, we remain with 7,061 observations. Table \ref{table:descriptives} presents some descriptive statistics for a selection of control variables. It shows that the group of individuals with and without health insurance coverage differ substantially. There are significant differences with respect to most of the control variables listed in the table. Females are significantly more likely to have health insurance coverage. Education and household income also show a significant positive correlation with health insurance coverage while the number of household members for example is negatively correlated with insurance coverage. Regarding the mediator, we find a similar pattern as for the treatment. With respect to many of the considered variables, the group of individuals who went for medical checkup differs substantially from those who did not. Further, we see that the correlation between many control variables and the treatment appear to have the same sign as that with the mediator.

\begin{table}[hp]
	\caption{Descriptive Statistics}\label{table:descriptives}
	{\footnotesize	
		\begin{center}
			\begin{tabular}{>{\raggedright\arraybackslash}p{4.10cm}lllllllll}
				\hline\hline
				& \textbf{overall} & \textbf{$D=1$} & \textbf{$D=0$} & \textbf{diff} & \textbf{p-val} & \textbf{$M=1$} & \textbf{$M=0$} & \textbf{diff} & \textbf{p-val} \\
				$n$ & 7,061 & 2,335 & 4,726 & &  & 3,612 & 3,449 &  &  \\
				\hline
				Female & 0.5 & 0.55 & 0.42 & 0.13 & 0 & 0.66 & 0.34 & 0.32 & 0 \\
				Age & 28.51 & 28.47 & 28.59 & -0.12 & 0 & 28.46 & 28.55 & -0.09 & 0.01 \\
				Ethnicity &&&&&&&&& \\
				\ \ \ \ \textit{Black} & 0.27 & 0.25 & 0.3 & -0.05 & 0 & 0.32 & 0.21 & 0.11 & 0 \\
				\ \ \ \ \textit{Hispanic} & 0.21 & 0.19 & 0.26 & -0.07 & 0 & 0.21 & 0.21 & 0 & 0.76 \\
				\ \ \ \ \textit{Mixed} & 0.01 & 0.01 & 0.01 & 0 & 0.34 & 0.01 & 0.01 & 0 & 0.42 \\
				\ \ \ \ \textit{White or Other} & 0.51 & 0.55 & 0.43 & 0.12 & 0 & 0.46 & 0.56 & -0.1 & 0 \\
				Relationship/Marriage &&&&&&&&& \\
				\ \ \ \ \textit{Not Cohabiting} & 0.62 & 0.61 & 0.65 & -0.03 & 0.01 & 0.61 & 0.64 & -0.03 & 0.01 \\
				\ \ \ \ \textit{Cohabiting} & 0.17 & 0.16 & 0.18 & -0.03 & 0.01 & 0.16 & 0.17 & 0 & 0.72 \\
				\ \ \ \ \textit{Married} & 0.18 & 0.21 & 0.14 & 0.07 & 0 & 0.2 & 0.17 & 0.03 & 0 \\
				\ \ \ \ \textit{Separated/ Widowed} & 0.02 & 0.02 & 0.03 & -0.01 & 0.03 & 0.02 & 0.02 & 0 & 0.62 \\
				\ \ \ \ \textit{Missing} & 0 & 0 & 0 & 0 & 0.5 & 0 & 0 & 0 & 0.79 \\
				Urban & 1.75 & 1.75 & 1.73 & 0.02 & 0.03 & 1.76 & 1.73 & 0.03 & 0.01 \\
				\ \ \ \ \textit{Missing} & 0.08 & 0.08 & 0.09 & -0.01 & 0.16 & 0.08 & 0.09 & -0.01 & 0.14 \\
				HH Income \tablefootnote{The HH income variable is the sum of several variables measuring HH income components (different sources and receivers). These variables are capped but only a total of 11 observations are in critical cap categories} & 43,406 & 48,388 & 33,322 & 15,066 & 0 & 44,217 & 42,556 & 1,661 & 0.24 \\
				\ \ \ \ \textit{Missing} & 0.21 & 0.19 & 0.24 & -0.05 & 0 & 0.2 & 0.22 & -0.01 & 0.2 \\
				HH Size & 3.09 & 3.06 & 3.15 & -0.1 & 0.04 & 3.13 & 3.05 & 0.08 & 0.07 \\
				\ \ \ \ \textit{Missing} & 0.06 & 0.05 & 0.07 & -0.02 & 0.01 & 0.05 & 0.07 & -0.01 & 0.03 \\
				HH Members under 18 & 0.69 & 0.65 & 0.76 & -0.11 & 0 & 0.77 & 0.6 & 0.17 & 0 \\
				\ \ \ \ \textit{Missing} & 0.06 & 0.06 & 0.07 & -0.02 & 0.01 & 0.06 & 0.07 & -0.01 & 0.04 \\
				Biological Children & 0.49 & 0.47 & 0.54 & -0.07 & 0 & 0.56 & 0.43 & 0.13 & 0 \\
				Highest Grade & 12.17 & 12.65 & 11.21 & 1.44 & 0 & 12.41 & 11.93 & 0.48 & 0 \\
				\ \ \ \ \textit{Missing} & 0.06 & 0.06 & 0.07 & -0.02 & 0 & 0.06 & 0.07 & -0.01 & 0.04 \\
				Employment &&&&&&&&& \\
				\ \ \ \ \textit{Employed} & 0.71 & 0.73 & 0.68 & 0.05 & 0 & 0.7 & 0.72 & -0.02 & 0.05 \\
				\ \ \ \ \textit{Unemployed} & 0.05 & 0.04 & 0.08 & -0.03 & 0 & 0.05 & 0.06 & -0.01 & 0.17 \\
				\ \ \ \ \textit{Out of Labor Force} & 0.21 & 0.19 & 0.24 & -0.04 & 0 & 0.21 & 0.2 & 0.01 & 0.4 \\
				\ \ \ \ \textit{Military} & 0.02 & 0.03 & 0.01 & 0.03 & 0 & 0.03 & 0.01 & 0.02 & 0 \\
				\ \ \ \ \textit{Missing} & 0 & 0 & 0 & 0 & 0.13 & 0 & 0 & 0 & 0.45 \\
				Working Hours {\tiny (per week)} & 24.83 & 25.47 & 23.53 & 1.94 & 0 & 24.44 & 25.24 & -0.81 & 0.09 \\
				\ \ \ \ \textit{Missing} & 0.06 & 0.05 & 0.07 & -0.02 & 0.01 & 0.05 & 0.07 & -0.01 & 0.04 \\
				Weight {\tiny (pounds)} & 157 & 157 & 157 & -1 & 0.64 & 154 & 160 & -6 & 0 \\
				\ \ \ \ \textit{Missing} & 0.08 & 0.08 & 0.1 & -0.02 & 0.01 & 0.08 & 0.09 & -0.01 & 0.09 \\
				Height {\tiny (feet)} & 5.12 & 5.18 & 5.01 & 0.17 & 0 & 5.08 & 5.17 & -0.09 & 0.02 \\
				\ \ \ \ \textit{Missing} & 0.09 & 0.08 & 0.11 & -0.04 & 0 & 0.08 & 0.09 & -0.01 & 0.13 \\
				Days 5+ drinks {\tiny (per month)} & 1.64 & 1.56 & 1.8 & -0.25 & 0.02 & 1.24 & 2.06 & -0.82 & 0 \\
				\ \ \ \ \textit{Missing} & 0.09 & 0.08 & 0.1 & -0.03 & 0 & 0.08 & 0.09 & -0.02 & 0.01 \\
				Days of Exercise {\tiny (per week)} & 2.39 & 2.41 & 2.36 & 0.05 & 0.42 & 2.32 & 2.46 & -0.14 & 0.01 \\
				\ \ \ \ \textit{Missing} & 0.05 & 0.05 & 0.06 & -0.01 & 0.03 & 0.04 & 0.06 & -0.01 & 0.03 \\
				Depressed/ Down  &&&&&&&&& \\
				\ \ \ \ \textit{Never} & 0.31 & 0.32 & 0.29 & 0.02 & 0.05 & 0.29 & 0.32 & -0.03 & 0.01 \\
				\ \ \ \ \textit{Sometimes} & 0.51 & 0.52 & 0.48 & 0.03 & 0.01 & 0.52 & 0.49 & 0.02 & 0.06 \\
				\ \ \ \ \textit{Mostly} & 0.09 & 0.09 & 0.1 & -0.01 & 0.14 & 0.1 & 0.08 & 0.02 & 0 \\
				\ \ \ \ \textit{Always} & 0.02 & 0.02 & 0.03 & -0.01 & 0 & 0.02 & 0.02 & 0 & 0.68 \\
				\ \ \ \ \textit{Missing} & 0.08 & 0.07 & 0.1 & -0.03 & 0 & 0.07 & 0.09 & -0.01 & 0.02 \\
				\hline
			\end{tabular}
		\end{center}
		\par
		Note: `overall', `$D=1$', `$D=0$', `$M=1$', `$M=0$' report the mean of the respective vaiable in the total sample, among treated, among non-treated, among mediated, and among non-mediated, respectively. `diff' and `p-val' provide the mean difference (across treatment or mediator states) and the p-value of a two-sample t-test, respectively.
	}
\end{table}

In order to assess the direct and indirect effect of health insurance coverage on general health, we consider estimation based on Theorem 1 and expression \eqref{1} derived by \cite{TchetgenTchetgenShpitser2011} well as (ii) Theorem 2 and expression \eqref{2}. We estimate the nuisance parameters and treatment effects in the same way as outlined in Section \ref{Sim} (i.e.\ post-lasso regression for modeling the nuisance parameters and 3-fold cross fitting for effect estimation) after augmenting the set of covariates with 3101 interaction and higher order terms. The  trimming threshold for discarding observations with too extreme propensity scores is set to $0.02$ (2\%), such that 893 and 136 observations are dropped when basing estimation on Theorems 1 and 2, respectively.

Table \ref{table:results} provides the estimated effects along with the standard error (`se') and p-value (`p-val') and also provides the estimated mean potential outcome under non-treatment for comparison (`$\hat{E}[Y(0,M(0))]$'). The ATEs of health insurance coverage on general health in the year 2008 (columns 2 and 8), estimated based on Theorems 1 or 2, are statistically significant at the 5\% and 10\% levels, respectively. As the outcome is measured on an ordinal scale ranging from `excellent' to `poor', the negative ATEs suggest a short-term health improving effect of health coverage. The direct effects under treatment (columns 3 and 9) under non-treatment (columns 4 and 10) are very similar to the ATEs and statistically significant (at least) at the 10\% level in 3 out ouf 4 cases. In contrast, the indirect effects under treatment (columns 5 and 11) and non-treatment (columns 6 and 12) are generally close to zero and not statistically significant at the 10\% level in 3 out of 4 cases. Thus, health insurance coverage does not seem to importantly affect general health of young adults in the U.S.\ through routine checkups in the short run, but rather through other mechanisms.

\begin{table}[H]
	\caption{Total, direct, and indirect effects on general health in 2008 }\label{table:results}
	{\footnotesize
		\begin{center}
			\begin{tabular}{r|cccccc|cccccc}
				\hline\hline
					& \multicolumn{6}{c|}{ Estimations based on Theorem 1  } & \multicolumn{6}{c}{ Estimations based on Theorem 2  } \\
				& $\hat{\Delta}$ & $\hat{\theta}(1)$ & $\hat{\theta}(0)$&  $\hat{\delta}(1)$ & $\hat{\delta}(0)$& $\hat{E}[Y(0,M(0))]$ & $\hat{\Delta}$ & $\hat{\theta}(1)$ & $\hat{\theta}(0)$&  $\hat{\delta}(1)$ & $\hat{\delta}(0)$& $\hat{E}[Y(0,M(0))]$ 	 \\
				\hline
	effect & -0.06 & -0.06 & -0.06 & -0.00 & 0.00 & 2.34  & -0.05 & -0.07 & -0.05 & 0.00 & 0.02 & 2.29 \\
	se & 0.03 & 0.04 & 0.03 & 0.01 & 0.02 & 0.02 & 0.03 & 0.03 & 0.03 & 0.01 & 0.01 & 0.03 \\
	p-val & 0.04 & 0.18 & 0.04 & 0.87 & 0.99 & 0.00 & 0.10 & 0.03 & 0.10 & 0.89 & 0.07 & 0.00 \\
				\hline
			\end{tabular}
		\end{center}
		\par
		Note: `effect', `se', and `p-val' report the respective effect estimate, standard error and p-value. Lasso regression is used for the estimation of nuisance parameters. The propensity score-based trimming threshold is set to $0.02$.
	}
\end{table}

\section{Conclusion}\label{conclusion}

In this paper, we combined causal mediation analysis with double machine learning under selection-on-observables assumptions which avoids adhoc pre-selection of control variables. Thus, this approach appears particularly fruitful in high-dimensional data with many potential control variables. We proposed estimators for natural direct and indirect effects as well as the controlled direct effect exploiting efficient score functions, sample splitting, and machine learning-based plug-in estimates for conditional outcome means, mediator densities, and/or treatment propensity scores. We demonstrated the root-$n$ consistency and asymptotic normality of the effect estimators under specific regularity conditions. Furthermore, we investigated the finite sample behavior of the proposed estimators in a simulation study and found the performance to be decent in samples with several thousand observations. Finally, we applied our method to data from the U.S.\ National Longitudinal Survey of Youth 1997 and found a moderate short-term effect of health insurance coverage on general health, which was, however, not mediated by routine checkups. The estimators considered in the simulation study and the application are available in the `causalweight' package for the statistical software `R'.

\bibliographystyle{econometrica}
\bibliography{research}

\bigskip

\renewcommand\appendix{\par
	\setcounter{section}{0}%
	\setcounter{subsection}{0}%
	\setcounter{table}{0}%
	\setcounter{figure}{0}%
	\renewcommand\thesection{\Alph{section}}%
	\renewcommand\thetable{\Alph{section}.\arabic{table}}}
\renewcommand\thefigure{\Alph{section}.\arabic{subsection}.\arabic{subsubsection}.\arabic{figure}}
\clearpage

\begin{appendix}
	
	\numberwithin{equation}{section}
	\noindent \textbf{\LARGE Appendices}

\section{Simulation results for standard errors}\label{simse}

\begin{table}[H]
	\caption{Simulation results for standard errors ($p=200$)}\label{table:sim1se}
	{\footnotesize
		\begin{center}
			\begin{tabular}{r|cccc|cccc|cccc|cccc}
				\hline\hline
				& \multicolumn{8}{c|}{ Coefficients given by $0.3/i^2$ for $i=1,...,p$ } & \multicolumn{8}{c}{Coefficients given by $0.5/i^2$ for $i=1,...,p$  }  \\
				& abias & sd & rmse & true & abias & sd & rmse & true & abias & sd & rmse & true & abias & sd & rmse & true \\
				& \multicolumn{4}{c|}{$n$=1000} & \multicolumn{4}{c|}{$n$=4000}   & \multicolumn{4}{c|}{$n$=1000} & \multicolumn{4}{c|}{$n$=4000}  \\
				\hline
					& \multicolumn{16}{c}{Double machine learning based on Theorem 1}   \\
				\hline
				    $se(\hat{\Delta})$ & 0.00 & 0.00 & 0.01 & 0.08 &  0.00 & 0.00 & 0.00 & 0.04 & 0.00 & 0.01 & 0.01 & 0.09 &  0.00 & 0.00 & 0.00 & 0.04 \\
					$se(\hat{\theta}(1))$  & 0.02 & 0.01 & 0.02 & 0.09 & 0.01 & 0.00 & 0.01 & 0.04 & 0.02 & 0.01 & 0.02 & 0.09 & 0.01 & 0.00 & 0.01 & 0.04 \\
					$se(\hat{\theta}(0))$ & 0.01 & 0.00 & 0.01 & 0.08 & 0.00 & 0.00 & 0.00 & 0.04 & 0.01 & 0.01 & 0.01 & 0.08 & 0.01 & 0.00 & 0.01 & 0.04 \\
					$se(\hat{\delta}(1))$ & 0.00 & 0.00 & 0.00 & 0.06 & 0.00 & 0.00 & 0.00 & 0.03 & 0.00 & 0.01 & 0.01 & 0.06 & 0.00 & 0.00 & 0.00 & 0.03 \\
					$se(\hat{\delta}(0))$ & 0.01 & 0.01 & 0.01 & 0.06 & 0.00 & 0.00 & 0.00 & 0.02 & 0.00 & 0.01 & 0.01 & 0.06 & 0.00 & 0.00 & 0.00 & 0.02 \\
						\hline
					& \multicolumn{16}{c}{Double machine learning based on Theorem 2}  \\
					\hline
					$se(\hat{\Delta})$ & 0.00 & 0.00 & 0.01 & 0.08 & 0.00 & 0.00 & 0.00 & 0.04 & 0.00 & 0.01 & 0.01 & 0.09 & 0.00 & 0.00 & 0.00 & 0.04 \\
					$se(\hat{\theta}(1))$  & 0.00 & 0.00 & 0.01 & 0.08 & 0.00 & 0.00 & 0.00 & 0.04 & 0.00 & 0.01 & 0.01 & 0.08 & 0.00 & 0.00 & 0.00 & 0.04 \\
					$se(\hat{\theta}(0))$ & 0.00 & 0.01 & 0.01 & 0.08 & 0.00 & 0.00 & 0.00 & 0.04 & 0.01 & 0.01 & 0.01 & 0.08 & 0.00 & 0.00 & 0.00 & 0.04 \\
					$se(\hat{\delta}(1))$ & 0.00 & 0.01 & 0.01 & 0.06 & 0.00 & 0.00 & 0.00 & 0.03 & 0.00 & 0.01 & 0.01 & 0.06 & 0.00 & 0.00 & 0.00 & 0.03 \\
					$se(\hat{\delta}(0))$ & 0.00 & 0.00 & 0.00 & 0.04 & 0.00 & 0.00 & 0.00 & 0.02 & 0.00 & 0.01 & 0.01 & 0.05 & 0.00 & 0.00 & 0.00 & 0.02 \\
				\hline
			\end{tabular}
		\end{center}
		\par
		Note: `abias', `sd', and `rmse' denote the absolute bias, standard deviation and root mean squared error of the respective standard error (`se').  `true' provides the true standard deviation.
	}
\end{table}

\newpage
\section{Proofs}

\small

For the proofs of Theorems 1 and 2, it suffices verifying the conditions of Assumptions 3.1 and 3.2 underlying Theorem 3.1 and 3.2 in \cite{Chetal2018}.

\subsection{Proof of Theorem 1} \label{proof1}

We first show that Assumptions 3.1 and 3.2 in \cite{Chetal2018} are satisfied for $\Psi_{d0}=E[Y(d,M(1-d))]$ based on \eqref{1}. Then, we show that Assumption 3.1 holds for $\Psi_{dm0}=E[Y(d,m)]$ based on \eqref{conddm}, but omit the proof of the validity of Assumption 3.2, as it follows in a very similar manner as for $\Psi_{d0}$. All bounds hold uniformly over all probability laws $P \in \mathcal{P}$, where $\mathcal{P}$ is the set of all possible probability laws, and we omit $P$ for brevity.

Let $\eta = (\mu(D,M,X),f(M|D,X), p_{d}(X))$  be the vector of nuisance parameters. Also, let $\mathcal{T}_n$ be the set of all $\eta=(\mu, f, p_d)$ in a neighbourhood of  $\eta_0$ that is shrinking with increasing $n,$ consisting of $P$-square integrable functions $\mu$, $f$, and $p_d$ such that
\begin{eqnarray} \label{Tn}
\left\|  \eta - \eta_0 \right\|_{q} &\leq& C,  \\
\left\|  \eta - \eta_0 \right\|_{2} &\leq& \delta_n, \notag \\
\left\|   p_d(X)-1/2\right\|_{\infty}  &\leq& 1/2-\epsilon, \notag \\
\left\|   f(M|D,X)-(\underline{f} + \overline{f})/2 \right\|_{\infty}  &\leq& (\overline{f} - \underline{f})/2, \notag \\
\left\|  \mu(D,M,X)-\mu_0(D,M,X)\right\|_{2} \times \left\|  p_{d}(X)-p_{d0}(X)\right\|_{2}  &\leq & \delta^{}_n n^{-1/2}, \notag \\
\left\|  \mu(D,M,X)-\mu_0(D,M,X)\right\|_{2} \times \left\|  f(M|1-D,X)-f_{0}(M|1-D,X)\right\|_{2}  &\leq & \delta^{}_n n^{-1/2}. \notag
\end{eqnarray}

We furthermore replace the sequence $(\delta_n)_{n \geq 1}$ by $(\delta_n')_{n \geq 1},$ where $\delta_n' = C_{\epsilon} \max(\delta_n,n^{-1/2}),$ where $C_{\epsilon}$ is sufficiently large constant that only depends on $C$ and $\epsilon.$ Let $R \equiv \overline{f}/\underline{f}$ stands for the maximal ratio of densities $f(m|d,X).$

\subsubsection{Counterfactual $E[Y(d,M(1-d))]$}  \label{Neyman}

The score function for the counterfactual $\Psi_{d0}=E[Y(d,M(1-d))]$ proposed by \cite{TchetgenTchetgenShpitser2011} is given by the following expression, with $W=(Y,M,D,X)$:
\begin{eqnarray}
\psi_d(W, \eta, \Psi_{d0}) &=& \frac{I\{D=d\} \cdot f(M|1-d,X)}{p_d(X)\cdot f(M|d,X) }\cdot[Y-\mu(d,M,X)] \notag\\
&& + \frac{I\{D=1-d\}}{1-p_d(X)}\cdot \Big[\mu(d,M,X) - \overbrace{\int_{m \in \mathcal{M}} \mu(d,m,X)\cdot f(m|1-d,X) dm}^{=: \nu(1-d,X)}  \Big]  \notag\\
&& + \underbrace{\int_{m \in \mathcal{M}} \mu(d,m,X) \cdot f(m|1-d,X) dm}_{=: \nu(1-d,X)} \ \ - \ \ \Psi_{d0}. \notag
\end{eqnarray}

\textbf{Assumption 3.1:  Moment Condition, Linear scores and Neyman orthogonality}
\vspace{5pt}

\textbf{Assumption 3.1(a)}

\textbf{Moment Condition:} The moment condition $E\Big[\psi_d(W, \eta_0, \Psi_{d0})\Big] = 0$ is satisfied:
\begin{eqnarray}
E\Big[\psi_d(W, \eta_0, \Psi_{d0})\Big] &=& E\Bigg[\overbrace{E\Bigg[\frac{I\{D=d\} \cdot f_0(M|1-d,X)}{p_{d0}(X)\cdot f_0(M|d,X) }\cdot [Y-\mu_0(d,M,X)]\Bigg|X\Bigg]}^{=E[E[Y-\mu_0(d,M,X)| D=d, M, X] |D=1-d,X] = 0} \Bigg] \notag\\
&& + \ E\Bigg[\overbrace{ E\Bigg[ \frac{I\{D=1-d\}}{1-p_{d0}(X)}\cdot [ \mu_0(d,M,X) -  \nu_0(1-d,X)] \Bigg| X \Bigg]}^{ =   E[\mu_0(d,M,X) -  \nu_0(1-d,X)|D=1-d,X] = 0   } \Bigg] \notag\\
&& + \  E[\nu_0(1-d,X)] \ \ - \ \ \Psi_{d0}  \notag\\
& = & \Psi_{d0}\ \ - \ \ \Psi_{d0} \ \ =  0, \notag
\end{eqnarray}
where the first equality follows from the law of iterated expectations. To better see this result, note that
\begin{eqnarray}
&&E\Bigg[\frac{I\{D=d\} \cdot f_0(M|1-d,X)}{p_{d0}(X)\cdot f_0(M|d,X) }\cdot [Y-\mu_0(d,M,X)]\Bigg|X\Bigg]\notag\\
&=&E\Bigg[ \frac{I\{D=d\} \cdot (1-p_{d0}(M,X)) }{ p_{d0}(M,X) \cdot (1-p_{d0}(X))}\cdot [Y-\mu_0(d,M,X)] \Bigg|X\Bigg]\notag\\
&=&E\Bigg[ E\Bigg[\frac{I\{D=d\}  }{ p_{d0}(M,X)  }\cdot [Y-\mu_0(d,M,X)]\Bigg|M,X\Bigg] \cdot  \frac{(1-p_{d0}(M,X))}{(1-p_{d0}(X))} \Bigg|X\Bigg]\notag\\
&=&E \Bigg[ E [Y-\mu_0(d,M,X) |D=d,M,X ] \cdot  \frac{(1-p_{d0}(M,X))}{(1-p_{d0}(X))}  \Bigg|X \Bigg]\notag\\
&=&E [ E [Y-\mu_0(d,M,X) |D=d,M,X ]   |D=1-d, X ]\notag\\
&=&E [ \mu_0(d,M,X)-\mu_0(d,M,X)   |D=1-d, X ]=0,\notag
\end{eqnarray}
where the first equality follows from Bayes' Law, the second from the law of iterated expectations, the third from basic probability theory, and the fourth from Bayes' Law. Furthermore, 
\begin{eqnarray}
&&E\Bigg[  \frac{I\{D=1-d\}}{1-p_{d0}(X)}\cdot [ \mu_0(d,M,X) -  \nu_0(1-d,X)]   \Bigg| X \Bigg]\notag\\
&=&E\Bigg[ E\Bigg[\frac{I\{D=1-d\}}{1-p_{d0}(X)}\cdot [ \mu_0(d,M,X) -  \nu_0(1-d,X)] \Big| M, X \Bigg] \Bigg| X \Bigg]\notag\\
&=&E\Bigg[    [ \mu_0(d,M,X) -  \nu_0(1-d,X)] \cdot \frac{1-p_{d0}(M,X)}{1-p_{d0}(X)}  \Bigg| X \Bigg]\notag\\
&=&E [     \mu_0(d,M,X) -  \nu_0(1-d,X)   | D=1-d, X ]=E [     \mu_0(d,M,X) | D=1-d, X ] -  \nu_0(1-d,X)  \notag\\
&=& \nu_0(1-d,X)  -  \nu_0(1-d,X) =0, \notag
\end{eqnarray}
where the first equality follows from the law of iterated expectations and the third from Bayes' Law.


\textbf{Assumption 3.1(b)}

\textbf{Linearity:} The score $ \psi_d(W, \eta_0, \Psi_{d0}) $ is linear in $ \Psi_{d0}$ as it can be written as:
 $\psi_d(W, \eta_0, \Psi_{d0}) = \psi_d^a(W, \eta_0) \cdot\Psi_{d0} + \psi_d^b(W, \eta_0) $
with $\psi_d^a(W, \eta_0) = -1$ and
{\small
\begin{eqnarray}
  \psi_d^b(W, \eta_0)& =& \frac{I\{D=d\} \cdot f_0(M|1-d,X)}{p_{d0}(X)\cdot f_0(M|d,X) }[Y-\mu_0(d,M,X)] \notag\\
  &+& \frac{I\{D=1-d\}}{1-p_{d0}(X)}\Big[\mu_0(d,M,X) - \nu_0(1-d,X) \Big] +  \nu_0(1-d,X). \notag
\end{eqnarray}
}

\textbf{Assumption 3.1(c)}

\textbf{Continuity:}
The expression for the second Gateaux derivative of a map $\eta \mapsto E\Big[\psi_d(W, \hat\eta, \Psi_{d0})\Big]$, given in \eqref{1}, is continuous.

\textbf{Assumption 3.1(d)}

\textbf{Neyman Orthogonality}: For any $\eta \in \mathcal{T}_n$, the Gateaux derivative in the direction $ \eta - \eta_0 =
(\mu(d,M,X) - \mu_0(d,M,X), f(M|D,X) - f_0(M|D,X), p_{d}(X) - p_{d0}(X)) $ is given by:

{\scriptsize
\begin{align}
&\partial E \big[\psi_d(W, \eta, \Psi_{d})\big] \big[\eta - \eta_0 \big] \notag\\
& = \resizebox{1.08\hsize}{!}{$E \Bigg[\frac{ \big[f(M|1-d,X) - f_0(M|1-d,X)\big] \cdot f_0(M|d,X) - \big[f(M|d,X) - f_0(M|d,X)\big]\cdot f_0(M|1-d,X)}{f_0^2(M|d,X)} \cdot \overbrace{\frac{I\{D=d\} }{p_{d0}(X)} \cdot  \Big(Y-\mu_0(d,M,X)\Big)}^{E[ \cdot|X]=E[Y-\mu_0(d,M,X)| D=d, X]  = 0}\Bigg]$} \notag\\
&\underbrace{ - \ E\Bigg[\underbrace{\frac{I\{D=1-d\}}{1-p_{d0}(X)}}_{E[ \cdot|X] = 1} \cdot \partial E[\nu_0(1-d,X)] [f(M|1-d,X) - f_0(M|1-d,X)] \Bigg] + \partial E[\nu_0(1-d,X)] [f(M|1-d,X) - f_0(M|1-d,X)] \Bigg] }_{ = 0}  \notag\\
&- \  E \Bigg[\underbrace{\frac{ I\{D=d\} \cdot f_0(M|1-d,X)}{p_{d0}(X) \cdot f_0(M|d,X)} \cdot \Big(Y-\mu_0(d,M,X)\Big)}_{ E[ \cdot|X]=E[E[Y-\mu_0(d,M,X)| D=d, M, X] |D=1-d,X] = 0} \cdot \frac{  p_{d}(X) - p_{d0}(X) }{p_{d0}(X)}\Bigg] \notag\\
& + \  E \Bigg[ \underbrace{\frac{I\{D=1-d\}}{(1-p_{d0}(X))} \cdot \Big(\mu_0(d,M,X) - \nu_0(1-d,X)\Big)}_{E[\cdot|X]=E[\mu_0(d,M,X) -  \nu_0(1-d,X)|D=1-d,X] = 0 }\cdot\frac{p_d(X) - p_{d0}(X)}{(1-p_{d0}(X))}\Bigg]  \notag \\
& - \ \underbrace{E \Bigg[\frac{I\{D=d\} \cdot f_0(M|1-d,X)}{p_{d0}(X) \cdot f_0(M|d,X)} \cdot \Big[\mu(d,M,X) - \mu_0(d,M,X)\Big]\Bigg]}_{E[\cdot] = E[E[ \cdot|M,X]]=  E\Big[\frac{p_{d0}(M,X) \cdot f_0(M|1-d,X)}{p_{d0}(X) \cdot f_0(M|d,X)} \cdot [\mu(d,M,X) - \mu_0(d,M,X)]\Big]}  \tag{$*$} \\
& + \ \underbrace{ E\Bigg[\frac{I\{D=1-d\}}{1-p_{d0}(X)} \cdot \Big[\mu(d,M,X) - \mu_0(d,M,X)\Big]\Bigg]}_{E[\cdot] = E[E[ \cdot|M,X]] = E\Big[ \frac{1-p_{d0}(M,X)}{1-p_{d0}(X)} \cdot [\mu(d,M,X) - \mu_0(d,M,X)] \Big]    } \tag{$**$} \\
& \underbrace{ - \ E\Bigg[\underbrace{\frac{I\{D=1-d\}}{1-p_{d0}(X)}\cdot \partial E[\nu_0(1-d,X)] [\mu(d,M,X) - \mu_0(d,M,X)]}_{E[ \cdot|X]= \frac{1-p_{d0}(X)}{1-p_{d0}(X)}\cdot \partial E[\nu_0(1-d,X)] [\mu(d,M,X) - \mu_0(d,M,X)]}  \Bigg]  + \partial E[\nu_0(1-d,X)] [\mu(d,M,X) - \mu_0(d,M,X)]}_{ = 0},  \notag
\end{align}
}
where terms $ (*) $ and  $ (**) $ cancel out by Bayes' Law, $\frac{p_{d0}(M,X)\cdot f_0(M|1-d,X)}{p_{d0}(X) \cdot f_0(M|d,X)}=\frac{p_{d0}(M,X)\cdot (1-p_{d0}(M,X))}{p_{d0}(M,X)\cdot (1-p_{d0}(X))}=\frac{1-p_{d0}(M,X)}{ 1-p_{d0}(X)}$.  Thus, it follows that:
\begin{align}
&\partial E \big[\psi_d(W, \eta, \Psi_{d})\big] \big[\eta - \eta_0 \big] = 0 \notag
\end{align}
proving that the score function is orthogonal.

\textbf{Assumption 3.1(e)}

\textbf{Singular values of $E[\psi_d^a(W;\eta_0)]$ are bounded:}
This holds trivially, because $\psi_d^a(W, \eta_0) = -1.$

\bigskip

\textbf{Assumption 3.2:  Score regularity and quality of nuisance parameter estimators}
\vspace{5pt}

\textbf{Assumption 3.2(a)}

This assumption follows directly from the regularity conditions (Assumption 4) and the definition of $\mathcal{T}_n$ given in (\ref{Tn}).
\\

\textbf{Assumption 3.2(b)}

\textbf{Bounds for $m_n$:}

We have
\begin{eqnarray}
\left\|  \mu_0(D,M,X)  \right\|_{q} &=& \left( E\left[ \left| \mu_0(D,M,X) \right|^q \right] \right)^{\frac{1}{q}} = \left( \sum_{d \in \{0,1\}} E\left[ \left|  \mu_0(d,M,X) \right|^q Pr  (D=d|M,X)  \right]  \right)^{\frac{1}{q}}\notag \\
&\geq& \epsilon^{1/q} \left( \sum_{d \in \{0,1\}} E\left[ \left|  \mu_0(d,M,X) \right|^q \right] \right)^{\frac{1}{q}} \notag \\
&\geq& \epsilon^{1/q} \left( \max_{d \in \{0,1\}} E\left[ \left|  \mu_0(d,M,X) \right|^q \right] \right)^{\frac{1}{q}} \notag \\
&=& \epsilon^{1/q}  \max_{d \in \{0,1\}} \left( E\left[ \left|  \mu_0(d,M,X) \right|^q \right] \right)^{\frac{1}{q}} =  \epsilon^{1/q}  \max_{d \in \{0,1\}} \left\|  \mu_0(d,M,X) \right\|_{q}. \notag
\end{eqnarray}
The first equality follows from definition, the second from the law of total probability, the first inequality from $Pr(D=d|M,X) \geq \epsilon.$
Using the same line of arguments we get that
$$\left\|  f_0(M|D,X)  \right\|_{q}  \geq   \epsilon^{1/q}  \max_{d \in \{0,1\}} \left\|  f_0(M|d,X) \right\|_{q} $$

Also, by Jensen's inequality $\left\|   \mu_0(D,M,X) \right\|_{q} \leq \left\|  Y  \right\|_{q}$, such that for any $d \in \{0,1\}$:
\begin{eqnarray}
 \left\|  \mu_0(d,M,X) \right\|_{q} &\leq& C/\epsilon^{1/q}, \label{mu} \\
 \left\|  f_0(M|d,X) \right\|_{q} &\leq& C/\epsilon^{1/q}, \notag
\end{eqnarray}
because of $\left \| Y \right \|_{q} \leq C$ by Assumption 4(a).

Similarly, for any $\eta \in \mathcal{T}_n$ we obtain:
\begin{eqnarray}
 \left\|  \mu(d,M,X) - \mu_0(d,M,X)  \right\|_{q} &\leq& C/\epsilon^{1/q}, \notag \\
 \left\|  f(M|d,X) - f_0(M|d,X)  \right\|_{q}&\leq& C/\epsilon^{1/q},  \notag
\end{eqnarray}
due to the definition of $\mathcal{T}_n$ given in (\ref{Tn}).

Also,
\begin{eqnarray} \label{nuDer}
\left\|  \nu_0(1-d,X)  \right\|_{q} &=& \left( E\left[ \left| \nu_0(1-d,X) \right|^q \right] \right)^{\frac{1}{q}} = \left( E\left[ \left| \int_{m \in \mathcal{M}} \mu_0(d,m,X)\cdot f_0(m|1-d,X) dm \right|^q \right] \right)^{\frac{1}{q}}  \\
&\leq& \left( E\left[ \int_{m \in \mathcal{M}}  \left| \mu_0(d,m,X) \right|^q \cdot f_0(m|1-d,X) dm  \right] \right)^{\frac{1}{q}}\notag  \\
&=& \left( E\left[ \int_{m \in \mathcal{M}}  \left| \mu_0(d,m,X) \right|^q \cdot f(m|d,X) \cdot \frac{f_0(m|1-d,X) }{f(m|d,X) }   dm  \right] \right)^{\frac{1}{q}} \notag \\
&\leq& R^{1/q} \left( E\left[ \int_{m \in \mathcal{M}}  \left| \mu_0(d,m,X) \right|^q \cdot f_0(m|d,X)  dm  \right] \right)^{\frac{1}{q}} \notag \\
&=& R^{1/q}  \left\|  \mu_0(d,M,X) \right\|_{q}  \leq \epsilon^{1/q} R^{1/q} \left\|  \mu_0(D,M,X)  \right\|_{q}  \notag
\end{eqnarray}
where we make use of the definition of $\nu_0$, Jensen's inequality, and the boundedness of the ratio of densities.  We therefore obtain $\left\|  \nu_0(1-d,X) \right\|_{q} \leq C/ (\epsilon^{1/q}R^{1/q})$ by inequality \eqref{mu}.

This permits bounding the following quantities:
\begin{eqnarray} \label{32b}
\left\|  \mu(d,M,X) \right\|_{q} &\leq& \left\|  \mu(d,M,X)-  \mu_0(d,M,X)  \right\|_{q} + \left\|  \mu_0(d,M,X)  \right\|_{q} \leq 2C/\epsilon^{1/q},  \\
\left\|  \nu(1-d,X)\right\|_{q} &\leq& \left\| \nu(1-d,X) -  \nu_0(1-d,X) \right\|_{q} + \left\| \nu_0(1-d,X) \right\|_{q} \leq 2C/ (\epsilon^{1/q}R^{1/q}), \notag \\
|\Psi_{d0} | &=& |E[  \nu_0(1-d,X) ] | \leq  E_{ } \Big[\left|   \nu_0(1-d,X)  \right|^1 \Big]^{\frac{1}{1}} =  \left\|  \nu_0(1-d,X)  \right\|_{1}  \notag \\
&\leq&  \left\| \nu_0(1-d,X) \right\|_{2} \leq  \left\| Y_2 \right\|_{2}/(\epsilon^{1/2} R^{1/2}) \overbrace{ \leq}^{q > 2}  \left\| Y_2 \right\|_{q}/(\epsilon^{1/2}R^{1/2}) \leq C / (\epsilon^{1/2}R^{1/2}),  \notag
\end{eqnarray}
using the triangular inequality, Jensen's inequality, and properties of statistical $l_q$ norms.

Finally, rearranging $\psi_d(W, \eta, \Psi_{d0})$
\begin{eqnarray}
\psi_d(W, \eta, \Psi_{d0})  &=&\underbrace{ \frac{I\{D=d\} \cdot f_0(M|1-d,X)}{p_{d}(X)\cdot f_0(M|d,X) } \cdot Y }_{=I_1}  \notag\\
& + & \underbrace{ \bigg( \frac{I\{D=1-d\}}{1-p_d(X)} -\frac{I\{D=d\} \cdot f_0(M|1-d,X)}{p_{d}(X)\cdot f_0(M|d,X) } \bigg) \cdot \mu(d,M,X)  }_{=I_2}  \notag\\
& + & \underbrace{  \bigg(1-   \frac{I\{D=1-d\}}{1-p_d(X)}  \bigg)  \nu(1-d,X)}_{=I_3} - \Psi_{d0},  \notag
\end{eqnarray}
provides
\begin{eqnarray}
\left\|  \psi_d(W, \eta, \Psi_{d0}) \right\|_{q} &\leq&  \left\| I_1 \right\|_{q}  + \left\| I_2 \right\|_{q}  + \left\| I_3 \right\|_{q} +  \left\| \Psi_{d0}   \right\|_{q} \notag \\
&\leq& \frac{R}{\epsilon}  \left\| Y \right\|_{q} \notag + \frac{1+R}{\epsilon} \left\|  \mu(d,M,X) \right\|_{q} + \\
&+& \frac{1-\epsilon}{\epsilon} \left\|  \nu(1-d,X)  \right\|_{q}   +  | \Psi_{d0} | \notag \\
&\leq& C \left( \frac{R}{\epsilon} + \frac{2}{\epsilon^{1+1/q}}\left( 1+R + \frac{1-\epsilon}{R^{1/q}} \right) + \frac{1}{\epsilon^{1/2}R^{1/2}} \right),  \notag
\end{eqnarray}
making use of the triangular inequality and inequalities (\ref{32b}). This provides the upper bound on $m_n$ in Assumption 3.2(b).

\textbf{Bound for $m'_n$:}

We note that
$$\Big(E[ |\psi_d^{a}(W, \eta) |^q] \Big)^{1/q}=1,$$
which provides the upper bound on $m'_n$ in Assumption 3.2(b).

\textbf{Assumption 3.2(c)}

\textbf{Bound for $r_n$:}

For any $\eta = (\mu,f,p_d)$ we have
$$ \Big| E\Big( \psi_d^{a}(W, \eta) - \psi_d^{a}(W, \eta_0) \Big) \Big| = |1-1| = 0 \leq \delta'_n,$$
providing the bound on $r_n$ in Assumption 3.2(c).

\textbf{Bound for $r'_n$:}

Using the triangular inequality
\begin{eqnarray*}
&&  \left\|   \psi_d(W, \eta, \Psi_{0d}) - \psi_d(W, \eta_0, \Psi_{0d}) \right\|_{2} \leq  \left\|    I\{D=d\} \cdot Y \cdot \left( \frac{f(M|1-d,X)}{p_d(X) f(M|d,X)} - \frac{f_0(M|1-d,X)}{p_{d0}(X) f_0(M|d,X)} \right)     \right\|_{2} \notag \\
&+& \left\|    I\{D=d\} \cdot \left( \frac{\mu(d,M,X) f(M|1-d,X)}{p_d(X) f(M|d,X)} - \frac{\mu_0(d,M,X) f_0(M|1-d,X)}{p_{d0}(X) f_0(M|d,X)} \right)     \right\|_{2} \notag \\
&+& \left\|    I\{D=1-d\} \cdot \left( \frac{\mu(d,M,X)}{1- p_d(X) } - \frac{\mu_0(d,M,X)}{1- p_{d0}(X) } \right)     \right\|_{2} +
 \left\|    I\{D=1-d\} \cdot \left( \frac{\nu(1-d,X)}{1- p_d(X) } - \frac{\nu_0(1-d,X)}{1- p_{d0}(X) } \right)     \right\|_{2} \notag \\
 &+&  \left\| \nu(1-d,X) - \nu_0(1-d,X) \right\|_{2} \notag \\
 &\leq& \delta_n \left( \frac{ C \cdot R^2}{\epsilon^2} +  \frac{C \cdot R^2}{\epsilon^2} \left(\frac{1}{\epsilon^{1/2}}+  \frac{C}{\epsilon^{1/2}} \right) +   \frac{1}{\epsilon^2} \left(\frac{1}{\epsilon^{1/2}} + \frac{C}{\epsilon^{1/2}} \right) +  \frac{1}{\epsilon^2} \left(\frac{1}{\epsilon^{1/2} R^{1/2}}  + \frac{C}{ R^{1/2}}  \right)  + \frac{1}{\epsilon^{1/2} R^{1/2}} \right) \leq \delta'_n,
\end{eqnarray*}
as long as $C_\epsilon$ in the definition of $\delta_n'$ is sufficiently large.  This gives the bound on $r'_n$ in Assumption 3.2(c). In order to show the second to the last inequalities, we provide bounds for the terms below, where we made use of the facts that $\left\| \mu(d,M,X) - \mu_0(d,M,X) \right\|_{2} \leq \delta_n/\epsilon^{1/2},$ and $\left\| \nu(1-d,X) - \nu_0(1-d,X) \right\|_{2} \leq \delta_n/(\epsilon^{1/2} R^{1/2})$ using similar steps as in Assumption 3.1(b) of \cite{Chetal2018}.

For the first term:
\begin{eqnarray*}
&& \left\| I\{D=d\} \cdot Y \cdot \left( \frac{f(M|1-d,X)}{p_d(X) f(M|d,X)} - \frac{f_0(M|1-d,X)}{p_{d0}(X) f_0(M|d,X)} \right)\right\|_{2} \leq C \cdot \left\|  \frac{f(M|1-d,X)}{p_d(X) f(M|d,X)} - \frac{f_0(M|1-d,X)}{p_{d0}(X) f_0(M|d,X)} \right\|_{2} \notag \\
&\leq& \frac{ C}{\epsilon^2 \underline{f}^2} \left\| f(M|1-d,X) f_0(M|1-d,X) p_{d0}(X) - f(M|1-d,X) f_0(M|1-d,X) p_{d}(X)  \right\|_{2} \notag \\
&\leq&  \frac{ C \cdot \overline{f}^2}{\epsilon^2 \underline{f}^2}  \left\| p_{d0}(X) - p_{d}(X) \right\|_{2} \leq  \delta_n \frac{ C \cdot R^2}{\epsilon^2},
\end{eqnarray*}
where we use $\left\| E[Y^2|d,M,X] \right\|_{\infty} \leq C^2$ (see our Assumption 4(a)) in the first inequality.

For the second term:
\begin{eqnarray*}
&&  \left\|    I\{D=d\} \cdot \left( \frac{\mu(d,M,X) f(M|1-d,X)}{p_d(X) f(M|d,X)} - \frac{\mu_0(d,M,X) f_0(M|1-d,X)}{p_{d0}(X) f_0(M|d,X)} \right)     \right\|_{2}  \notag \\
&\leq& \left\|  \frac{\mu(d,M,X) f(M|1-d,X)}{p_d(X) f(M|d,X)} - \frac{\mu_0(d,M,X) f_0(M|1-d,X)}{p_{d0}(X) f_0(M|d,X)}    \right\|_{2}  \notag \\
&\leq& \frac{C}{\epsilon^2 \underline{f}^2} \left\| \mu(d,M,X) f(M|1-d,X) f_0(M|1-d,X) p_{d0}(X) -  \mu_0(d,M,X) f(M|1-d,X) f_0(M|1-d,X) p_{d}(X)  \right\|_{2} \notag \\
&\leq&  \frac{C \overline{f}^2}{\epsilon^2 \underline{f}^2}  \left\| \mu(d,M,X) p_{d0}(X) - \mu_0(d,M,X) p_{d}(X) \right\|_{2} \notag \\
&=& \frac{C \cdot R^2}{\epsilon^2} \left\| \mu(d,M,X) p_{d0}(X) - \mu_0(d,M,X) p_{d}(X) + \mu_0(d,M,X) p_{d0}(X)  - \mu_0(d,M,X) p_{d0}(X)  \right\|_{2} \notag \\
&\leq& \frac{C \cdot R^2}{\epsilon^2}\big( \left\| p_{d0}(X) (\mu(d,M,X)  - \mu_0(d,M,X))  \right\|_{2} +  \left\| \mu_0(d,M,X) (p_{d0}(X)  -  p_{d}(X))  \right\|_{2}  \big) \notag \\
&\leq& \frac{C \cdot R^2}{\epsilon^2}\big( \left\| \mu(d,M,X)  - \mu_0(d,M,X)  \right\|_{2} +  C\left\| (p_{d0}(X)  -  p_{d}(X))  \right\|_{2}  \big) \notag \\
&\leq& \frac{C \cdot R^2}{\epsilon^2} \left(\frac{\delta_n}{\epsilon^{1/2}}+  C\delta_n \right) = \delta_n \frac{C \cdot R^2}{\epsilon^2} \left(\frac{1}{\epsilon^{1/2}}+  C \right)
 \notag
\end{eqnarray*}
where the fifth inequality follows from $E[Y^2| D=d,M,X] \geq (E[Y| D=d, M, X])^2= \mu^2_0(d,M,X) $ by conditional Jensen's inequality and therefore $\left\| \mu_0 (d,M,X)
\right\|_\infty \leq C^2.$

For the third term:
\begin{eqnarray*}
&& \left\|    I\{D=1-d\} \cdot \left( \frac{\mu(d,M,X)}{1- p_d(X) } - \frac{\mu_0(d,M,X)}{1- p_{d0}(X) } \right)     \right\|_{2} \leq  \left\|  \frac{\mu(d,M,X)}{1- p_d(X) } - \frac{\mu_0(d,M,X)}{1- p_{d0}(X) }    \right\|_{2} \notag \\
&\leq& \frac{1}{\epsilon^2} \left\| \mu(d,M,X)p_{1-d,0} - \mu_0(d,M,X)p_{1-d}     \right\|_{2}  \notag \\
&=& \frac{1}{\epsilon^2} \left\| \mu(d,M,X)p_{1-d,0} - \mu_0(d,M,X)p_{1-d} + \mu_0(d,M,X)p_{1-d,0} - \mu_0(d,M,X)p_{1-d,0}     \right\|_{2} \notag \\
&\leq& \frac{1}{\epsilon^2}  \big(  \left\| p_{1-d,0} (\mu(d,M,X) - \mu_0(d,M,X))   \right\|_{2} +  \left\|  \mu_0(d,M,X)( p_{1-d,0} - p_{1-d})     \right\|_{2} \big) \notag \\
&\leq& \frac{1}{\epsilon^2}  \big(  \left\|  \mu(d,M,X) - \mu_0(d,M,X)   \right\|_{2} + C \left\| p_{1-d,0} - p_{1-d}     \right\|_{2} \big) \notag \\
&\leq& \frac{1}{\epsilon^2} \left(\frac{\delta_n}{\epsilon^{1/2}} + C \delta_n \right) =  \delta_n \frac{1}{\epsilon^2} \left(\frac{1}{\epsilon^{1/2}} + C \right).
\end{eqnarray*}

For the fourth term:
\begin{eqnarray*}
&& \left\|    I\{D=1-d\} \cdot \left( \frac{\nu(1-d,X)}{1- p_d(X) } - \frac{\nu_0(1-d,X)}{1- p_{d0}(X) } \right)     \right\|_{2} \notag \\
&\leq& \frac{1}{\epsilon^2}  \big(  \left\| p_{1-d,0} (\nu(1-d,X) - \nu_0(1-d,X))   \right\|_{2} +  \left\|  \nu_0(1-d,X)( p_{1-d,0} - p_{1-d})     \right\|_{2} \big) \notag \\
&\leq& \frac{1}{\epsilon^2}  \big(  \left\|  \nu(1-d,X) - \nu_0(1-d,X)   \right\|_{2} + \frac{C}{ R^{1/2}}  \left\| p_{1-d,0} - p_{1-d}     \right\|_{2} \big) \notag \\
&\leq& \frac{1}{\epsilon^2} \left(\frac{\delta_n}{\epsilon^{1/2} R^{1/2}}  + \frac{C}{ R^{1/2}}  \delta_n \right) = \delta_n \frac{1}{\epsilon^2} \left(\frac{1}{\epsilon^{1/2} R^{1/2}}  + \frac{C}{R^{1/2}}  \right),
\end{eqnarray*}
where we used Jensen's inequality similarly to \ref{nuDer} in order to get $E[\nu^2_0(1-d,X)] \leq R \cdot E[\mu_0^2(d,M,X)]$ and hence$\left\| \nu_0 (1-d,X) \right\|_\infty \leq C^2/R$.


\textbf{Bound on $\lambda'_n$:}
Consider
\begin{equation}
f(r) := E[\psi(W,\eta_0 + r(\eta-\eta_0),\Psi_{d0})] \notag
\end{equation}

We subsequently omit arguments for the sake of brevity and use $p_{d}= p_{d}(X),f_d = f_d(M|d,X), \mu = \mu(d,M,X),  \nu = \nu(1-d,X)$ and similarly $p_{d0}, f_{0d}, \mu_0, \nu_0.$

 For any $r \in (0,1):$
\begin{eqnarray} \label{secondGatDer}
\frac{\partial^2 f(r)}{\partial r^2}&=& E\Bigg[  2\cdot I\{D = 1-d\} \frac{ (\mu-\mu_0)(p_{d} - p_{d0})}{\left(1 - p_{d0} + r(p_{d0} -p_{d})\right)^2}  \Bigg] + E\Bigg[  2\cdot I\{D = 1-d\} \frac{(\nu-\nu_0)(p_{d} - p_{d0})}{\left(1 - p_{d0} + r(p_{d0} -p_{d})\right)^2}  \Bigg] \notag \\ 
&+& E\Bigg[  2 \cdot I\{D = d\} \frac{(f_{d} -f_{d0})(f_{1-d} -f_{1-d,0}) \left(Y-\mu_0 - r(\mu-\mu_0) \right)}{\left(p_{d0} + r(p_{d} -p_{d0})\right) \left(f_{d0} + r(f_{d} -f_{d0})\right)^2}  \Bigg] \notag \\ 
&+& E\Bigg[  2\cdot I\{D = d\} \frac{(p_{d} -p_{d0})(f_{1-d} -f_{1-d,0}) \left(Y-\mu_0 - r(\mu-\mu_0) \right)}{\left(p_{d0} + r(p_{d} -p_{d0})\right)^2 \left(f_{d0} + r(f_{d} -f_{d0})\right)}  \Bigg] \notag \\ 
&+& E\Bigg[  2\cdot I\{D = d\} \frac{(f_{d} -f_{d0})\left(f_{1-d,0} + r(f_{1-d} -f_{1-d,0})\right) \left( - (\mu-\mu_0) \right)}{\left(p_{d0} + r(p_{d} -p_{d0})\right) \left(f_{d0} + r(f_{d} -f_{d0})\right)^2}  \Bigg] \notag \\ 
&+& E\Bigg[  2 \cdot I\{D = d\} \frac{(p_{d} -p_{d0})\left(f_{1-d,0} + r(f_{1-d} -f_{1-d,0})\right)  \left(-(\mu-\mu_0) \right)}{\left(p_{d0} + r(p_{d} -p_{d0})\right)^2 \left(f_{d0} + r(f_{d} -f_{d0})\right)}  \Bigg] \notag \\  
&+& E\Bigg[  (-2) \cdot I\{D = d\} \frac{\left(f_{1-d} -f_{1-d,0}\right)  (\mu-\mu_0)}{\left(p_{d0} + r(p_{d} -p_{d0})\right) \left(f_{d0} + r(f_{d} -f_{d0})\right)}  \Bigg] \notag \\  
&+& E\Bigg[  2\cdot I\{D = d\} \frac{(f_{d} -f_{d0})^2\left(f_{1-d,0} + r(f_{1-d} -f_{1-d,0})\right) \left(Y-\mu_0 - r(\mu-\mu_0) \right)}{\left(p_{d0} + r(p_{d} -p_{d0})\right) \left(f_{d0} + r(f_{d} -f_{d0})\right)^3}  \Bigg] \notag \\  
&+& E\Bigg[  2\cdot I\{D = d\} \frac{(f_{d} -f_{d0}) (p_{d} -p_{d0}) \left(f_{1-d,0} + r(f_{1-d} -f_{1-d,0})\right) \left(Y-\mu_0 - r(\mu-\mu_0) \right)}{\left(p_{d0} + r(p_{d} -p_{d0})\right)^2 \left(f_{d0} + r(f_{d} -f_{d0})\right)^2}  \Bigg] \notag \\  
&+& E\Bigg[  2\cdot I\{D = d\} \frac{(p_{d} -p_{d0})^2\left(f_{1-d,0} + r(f_{1-d} -f_{1-d,0})\right) \left(Y-\mu_0 - r(\mu-\mu_0) \right)}{\left(p_{d0} + r(p_{d} -p_{d0})\right)^3 \left(f_{d0} + r(f_{d} -f_{d0})\right)}  \Bigg] \notag \\ 
&+& E\Bigg[  2\cdot I\{D = 1-d\} \frac{\left(\mu_0 - \nu_0  \right) (p_{d} - p_{d0})^2}{\left(1 - p_{d0} + r(p_{d0} -p_{d})\right)^3}  \Bigg]  \notag \\ 
&+& E\Bigg[  2\cdot I\{D = 1-d\} \frac{\left(r(\mu-\mu_0) -  r(\nu-\nu_0) \right) (p_{d} - p_{d0})^2}{\left(1 - p_{d0} + r(p_{d0} -p_{d})\right)^3}  \Bigg]  \notag 
\end{eqnarray}

Note that the following inequalities can be shown to hold using similar steps as in Assumption 3.1(b) of \cite{Chetal2018}:
\begin{eqnarray}
\left\| \mu -\mu_0\right\|_{2} &=& \left\| \mu(d,M,X) - \mu_0(d,M,X)  \right\|_{2} \leq  \left\|   \mu(D,M,X) - \mu_0(D,M,X)  \right\|_{2}/\epsilon^{1/2} \leq \delta_n/\epsilon^{1/2}, \notag  \\
\left\| \nu -\nu_0\right\|_{2} &=& \left\| \nu(1-d,X) - \nu_0(1-d,X)  \right\|_{2} \leq  \left\|   \mu(D,M,X) - \mu_0(D,M,X)  \right\|_{2}/\epsilon^{1/2}R^{1/2} \leq  \delta_n/(\epsilon^{1/2} R^{1/2}), \notag
\end{eqnarray}

These inequalities together with our Assumption 4 imply
\begin{eqnarray}
E[Y-\mu_0(d,M,X)|D=d,M,X]  &= & 0, \notag \\
|p_{d} - p_{d0}|  &\leq&  2, \notag \\
\left\| \mu_0 \right\|_{q} \leq  \left\| Y \right\|_{q}/\epsilon^{1/q} &\leq & C/\epsilon^{1/q} \notag \\
\left\|  \mu-\mu_0\right\|_{2} \times \left\|  p_{d}-p_{d0}\right\|_{2} &\leq & \delta^{}_n n^{-1/2}/\epsilon^{1/2},\notag \\
\left\|  \mu-\mu_0\right\|_{2} \times \left\|  f_{1-d}-f_{1-d,0}\right\|_{2} &\leq & \delta^{}_n n^{-1/2}/(\epsilon R^{1/2}),\notag
\end{eqnarray}
for all $d \in \{1,0\}$ and consequently
$$\left\|  \nu-\nu_0\right\|_{2} \times \left\|  p_{d}-p_{d0}\right\|_{2} \leq  \delta^{}_n n^{-1/2}/(\epsilon^{1/2} R^{1/2}).$$

Putting everything together, we get that for some value $C_{\epsilon}''$ that only depends on $C$ and $\epsilon$
\begin{equation}
\left|\frac{\partial^2 f(r)}{\partial r^2} \right| \leq C_{\epsilon}'' \delta_n n^{-1/2} \leq \delta_n' n^{-1/2}. \notag
\end{equation}
This gives the upper bound on $\lambda'_n$ in Assumption 3.2(c) as long as $C_{\epsilon}$ in the definition of $\delta'_n$ satisfies $C_{\epsilon} \geq C_{\epsilon}''$.

In order to verify that this inequality holds we consider all the terms in $\frac{\partial^2 f(r)}{\partial r^2}$ separately.  For the first term we obtain
\begin{eqnarray*}
&&\left|  E\Bigg[  2\cdot I\{D = 1-d\} \frac{ (\mu-\mu_0)(p_{d} - p_{d0})}{\left(1 - p_{d0} + r(p_{d0} -p_{d})\right)^2}  \Bigg] \right| \leq  \frac{2}{\epsilon^3}\left|  E\Bigg[ (\mu-\mu_0)(p_{d} - p_{d0})  \Bigg] \right| \leq  \frac{2}{\epsilon^3} \frac{\delta_n}{\epsilon^{1/2}R^{1/2}} n^{-1/2}, \notag \\
\end{eqnarray*}
where we made use of the fact that $1 \geq p_{d0} + r(p_{d} -p_{d0}) =  (1-r)p_{d0} + r p_{d} \geq (1-r)\epsilon + r \epsilon = \epsilon,$  $\underline{f} \leq f_{d0} + r(f_{d} -f_{d0}) \leq \overline{f} $, and Holder's inequality. For the third term we obtain
\begin{eqnarray*}
&&\left| E\Bigg[  2 \cdot I\{D = d\} \frac{(f_{d} -f_{d0})(f_{1-d} -f_{1-d,0}) \left(Y-\mu_0 - r(\mu-\mu_0) \right)}{\left(p_{d0} + r(p_{d} -p_{d0})\right) \left(f_{d0} + r(f_{d} -f_{d0})\right)^2}  \Bigg] \right| \notag \\
&\leq& \frac{2}{\epsilon \underline{f}^2}  (\overline{f} - \underline{f})^2 \left|  E\Bigg[ I\{D = d\}  \left(Y-\mu_0 \right) \Bigg] \right| + \frac{2}{\epsilon \underline{f}^2}  \left|  E\Bigg[ I\{D = d\} (f_{d} -f_{d0})(f_{1-d} -f_{1-d,0}) r(\mu-\mu_0) \Bigg] \right| \notag \\
&\leq&  \frac{2}{\epsilon \underline{f}^2} (\overline{f} - \underline{f})  \left|  E\Bigg[ 1 \cdot (f_{1-d} -f_{1-d,0}) (\mu-\mu_0) \Bigg] \right| \leq \frac{2}{\epsilon \underline{f}^2} (\overline{f} - \underline{f}) \frac{\delta_n}{\epsilon^{1/2}} n^{-1/2}.
\end{eqnarray*}

And for the second to the last terms, we obtain
\begin{eqnarray*}
&&\left|  E\Bigg[  2\cdot I\{D = 1-d\} \frac{\left(\mu_0 - \nu_0  \right) (p_{d} - p_{d0})^2}{\left(1 - p_{d0} + r(p_{d0} -p_{d})\right)^3}  \Bigg]  \right| \notag \\
 &=& \left|E\Bigg[ \overbrace{ I\{D=1-d\} \frac{\left(\mu_0 - \nu_0  \right) }{p_{1-d,0}} }^{E[E[\cdot|M,X]|X]=0 } \cdot \frac{ p_{1-d,0}(p_{d} - p_{d0})^2}{\left(1 - p_{d0} + r(p_{d0} -p_{d})\right)^3}  \Bigg]  \right|  = 0. \notag
\end{eqnarray*}
All the remaining terms are bounded similarly.

\textbf{Assumption 3.2(d)}

Finally, we consider
\begin{eqnarray}
E\Big[ ( \psi_d(W, \eta, \Psi_{d0}) )^2\Big]  &=& E\Bigg[ \Bigg( \underbrace{ \frac{I\{D=d\} \cdot f_0(M|1-d,X)}{p_{d}(X)\cdot f_0(M|d,X) } \cdot \left( Y -\mu_0(d,M,X) \right) }_{=I_1}  \notag\\
& + & \underbrace{ \bigg( \frac{I\{D=1-d\}}{1-p_d(X)} \bigg) \cdot \left( \mu_0(d,M,X) -  \nu_0(1-d,X) \right) }_{=I_2}   +  \underbrace{  \nu_0(1-d,X) - \Psi_{d0} }_{=I_3} \Bigg)^2  \Bigg]   \notag\\
&=&  E[I_1^2 + I_2^2 + I_3^2] \geq E[I^2_1]\notag\\
& = & E\Bigg[  \Bigg( \frac{I\{D=d\} \cdot f_0(M|1-d,X)}{p_{d}(X)\cdot f_0(M|d,X) } \Bigg)^2 \left( Y -\mu_0(d,M,X) \right)^2 \Bigg] \notag\\
& \geq & \frac{\underline{f}^2}{(1-\epsilon)\overline{f}^2} E\Bigg[ \left( Y -\mu_0(d,M,X) \right)^2 \Bigg]  \geq \frac{c^2}{(1-\epsilon)R^2} > 0, \notag
\end{eqnarray}
where the second equality follows from
\begin{eqnarray}
E\Big[ I_1 \cdot I_2\Big] &=& E\Bigg[  \overbrace{\frac{I\{D=d\} \cdot f_0(M|1-d,X)}{p_{d}(X)\cdot f_0(M|d,X) }  \frac{I\{D=1-d\}}{1-p_d(X)}}^{I\{D=d\}\cdot I\{D=1-d\} = 0 }  \cdot \left( Y -\mu_0(d,M,X) \right)  \cdot  \left( \mu_0(d,M,X) -  \nu_0(1-d,X) \right) \Bigg], \notag \\
E\Big[ I_2 \cdot I_3\Big] &=& E\Bigg[ \overbrace{  \frac{I\{D=1-d\}}{1-p_d(X)} \cdot \left( \mu_0(d,M,X) -  \nu_0(1-d,X) \right)  }^{E[\cdot|X]=0}  \cdot (\nu_0(1-d,X) - \Psi_{d0}) \Bigg],\notag\\
E\Big[ I_1 \cdot I_3\Big] &=& E\Bigg[  \overbrace{\frac{I\{D=d\} \cdot f_0(M|1-d,X)}{p_{d}(X)\cdot f_0(M|d,X) } \cdot   \left( Y -\mu_0(d,M,X) \right) }^{E[ \cdot|X]= 0}   \cdot   (\nu_0(1-d,X) - \Psi_{d0})  \Bigg]. \notag
\end{eqnarray}

\subsubsection{Counterfactual $E[Y(d,m)]$}  \label{conscoreproof}

The score for the estimation of $E[Y(d,m)]$ based on (\ref{conddm}) is given by:
\begin{align}
E \big[\psi_{dm}(W, \eta, \Psi_{dm0})\big] =  \   & E\Bigg[\frac{I\{D=d\}\cdot I\{M=m\}\cdot[Y-\mu(d,m,X)]}{ f(m|d,X)\cdot p_d(X)} + \mu(d,m,X)-\Psi_{dm0} \Bigg]. \notag
\end{align}

\textbf{Assumption 3.1:  Moment Condition, Linear scores and Neyman orthogonality}
\vspace{5pt}

\textbf{Assumption 3.1(a)}

\textbf{Moment condition:} The moment condition $E\Big[\psi_{dm}(W, \eta_0, \Psi_{dm0})\Big] = 0$ is satisfied:
\begin{align}
E\Big[\psi_{dm}(W,  \eta_0, \Psi_{dm0})\Big]  = & \ E\Bigg[\frac{I\{D=d\}\cdot I\{M=m\}\cdot[Y- \mu_0(d,m,X)]}{  f_0(m|d,X)\cdot p_{d0}(X)} +  \mu_0(d,m,X)-\Psi_{dm0} \Bigg] \notag \\
= \  &  E\Bigg[\overbrace{E\Bigg[\frac{I\{D=d\}\cdot I\{M=m\}\cdot [Y- \mu_0(d,m,X)]} { f_0(m|d,X)\cdot  p_{d0}(X)}\Bigg| X\Bigg]}^{=E[Y- \mu_0(d,m,X)| d,m,X]=0}\Bigg] + E\Big[  \mu_0(d,m,X)\Big] -\Psi_{dm0} \notag \\
=  \   &  \Psi_{dm0}-\Psi_{dm0}=0. \notag
\end{align}

\textbf{Assumption 3.1(b)}

\textbf{Linearity:} The score $ \psi_{dm}(W, \eta_0, \Psi_{dm0}) $ is linear in $ \Psi_{dm0} $ as it can be written as:
$\psi_{dm}(W, \eta_0, \Psi_{dm0}) = \psi_d^a(W, \eta_0) \cdot \Psi_{dm0} + \psi_d^b(W, \eta_0) $
with $\psi_d^a(W, \eta_0) = -1$ and
\small
\begin{eqnarray}
\psi_d^b(W, \eta_0) = \frac{I\{D=d\}\cdot I\{M=m\}\cdot[Y-\mu(d,m,X)]}{ f(m|d,X)\cdot p_d(X)} + \mu(d,m,X) \notag
\end{eqnarray}

\textbf{Assumption 3.1(c)}
\textbf{Continuity:}
The expression for the second Gateaux derivative of a map $\eta \mapsto E\Big[\psi_{dm}(W, \eta, \Psi_{dm0})\Big]$, is continuous.
\bigskip
\\

\textbf{Assumption 3.1(d)}

\textbf{Neyman orthogonality}: The Gateaux derivative in the direction $ \eta - \eta_0 =
( \mu(d,M,X) - \mu_0(d,M,X), f(M|D,X) - f_0(M|D,X), p_{d}(X) - p_{d0}(X)) $ is given by:
\begin{align}
\partial E & \big[\psi_{dm}(W, \eta, \Psi_{dm})\big] \big[\eta - \eta_0 \big]\notag \\
= & \overbrace{-E\Bigg[\underbrace{\frac{I\{D=d\}\cdot I\{M=m\}}{ f_0(m|d,X)\cdot p_{d0}(X)}}_{E[ \cdot | X] = \frac{\Pr(D=d,M=m|X)}{\Pr(D=d,M=m|X)}=1}\cdot \Big[\mu(d,m,X) - \mu_0(d,m,X)\Big]\Bigg]+E\Big[\mu(d,m,X) - \mu_0(d,m,X)\Big]}^{=0}\notag\\
& -E\Bigg[\frac{\overbrace{I\{D=d\}\cdot I\{M=m\}\cdot[Y-\mu_0(d,m,X)]}^{E[ \cdot | X] = E[Y- \mu_0(d,m,X)| d,m,X]=0}}{f_0(m|d,X)\cdot p_{d0}(X)}\cdot \frac{ f(m|d,X)-f_0(m|d,X) }{ f_0(m|d,X)}\Bigg]\notag\\
& -E\Bigg[\frac{\overbrace{I\{D=d\}\cdot I\{M=m\}\cdot[Y-\mu_0(d,m,X)]}^{E[ \cdot | X] = E[Y- \mu_0(d,m,X)| d,m,X]=0}}{f_0(m|d,X)\cdot p_{d0}(X)}\cdot \frac{ p_d(X)-p_{d0}(X) }{p_{d0}(X)}\Bigg] .\notag
\end{align}
Thus, it follows that:
\begin{align}
&\partial E \big[\psi_{dm}(W, \eta, \Psi_{dm})\big] \big[\eta - \eta_0 \big] = 0 \notag
\end{align}
proving that the score function is orthogonal.

\textbf{Assumption 3.1(e)}

\textbf{Singular values of $E[\psi^{a}_{d}(W;\eta_0)]$ are bounded:}
This holds trivially, because $\psi^{a}_{d}(W;\eta_0) = -1.$

\textbf{Assumption 3.2:  Score regularity and quality of nuisance parameter estimators}
\vspace{5pt}
This proof is omitted for the sake of brevity. It follows along similar lines as the proof for $Y(d,M(1-d))$ presented in subsection \ref{Neyman}.

This concludes the proof of Theorem 1. $\hfill\square$

\subsection{Proof of Theorem 2} \label{altscoreproof}

The alternative score for the counterfactual based on \eqref{2} is given by:
\begin{align}
\psi_d^*(W, \eta^*, \Psi_{d0}) &= &  E\Bigg[\frac{I\{D=d\} \cdot (1-p_d(M,X))}{p_d(M,X) \cdot (1-p_d(X))} \cdot \Big[Y-\mu(d,M,X)\Big] \notag  \\
& +& \frac{I\{D=1-d\}}{1-p_d(X)} \cdot \Bigg[\mu(d,M,X) - \overbrace{E\Big[\mu(d,M,X) \Big|  D=1-d, X\Big]}^{=: \omega(1-d,X)}\Bigg] \notag \\
& + &\overbrace{E\Big[\mu(d,M,X)  \Big| D=1-d, X\Big]}^{=: \omega(1-d,X)}\Bigg] -\Psi_{d0} \notag
\end{align}
with $ \eta^* = (\mu(D,M,X),\omega(D,X),p_d(M,X),p_d(X)) $.

Let $\mathcal{T}^*_n$ be the set of all $\eta^*$ consisting of $P$-square integrable functions $\mu(D,M,X),\omega(D,X),p_d(M,X)$, and $p_d(X)$ such that
\begin{eqnarray} \label{Tnstar}
\left\|  \eta^* - \eta^*_0 \right\|_{q} &\leq& C,  \\
\left\|  \eta^* - \eta^*_0 \right\|_{2} &\leq& \delta_n, \notag \\
\left\|   p_d(X)-1/2\right\|_{\infty}  &\leq& 1/2-\epsilon, \notag \\
\left\|   p_{d}(M,X)-1/2\right\|_{\infty}  &\leq& 1/2-\epsilon, \notag \\
\left\|  \mu(D,M,X)-\mu_0(D,M,X)\right\|_{2} \times \left\|  p_{d}(X)-p_{d0}(X)\right\|_{2}  &\leq & \delta^{}_n n^{-1/2}, \notag\\
\left\|  \mu(D,M,X)-\mu_0(D,M,X)\right\|_{2} \times \left\|  p_{d}(M,X)-p_{d0}(M,X)\right\|_{2}  &\leq & \delta^{}_n n^{-1/2}, \notag\\
\left\|  \omega(D,M,X)-\omega_0(D,M,X)\right\|_{2} \times \left\|  p_{d}(X)-p_{d0}(X)\right\|_{2}  &\leq & \delta^{}_n n^{-1/2}. \notag
\end{eqnarray}

We replace the sequence $(\delta_n)_{n \geq 1}$ by $(\delta_n')_{n \geq 1},$ where $\delta_n' = C_{\epsilon} \max(\delta_n,n^{-1/2}),$ where $C_{\epsilon}$ is a sufficiently large value that only depends on $C$ and $\epsilon.$

\textbf{Assumption 3.1:  Moment Condition, Linear scores and Neyman orthogonality}
\vspace{5pt}

\textbf{Assumption 3.1(a)}

\textbf{Moment condition:} The moment condition $E\Big[\psi_d^*(W, \eta_0^*, \Psi_{d0})\Big] = 0$ is satisfied:
\begin{eqnarray}
E\Big[\psi_d^*(W,  \eta_0^*, \Psi_{d0})\Big] &=& E\Bigg[ \overbrace{E\Bigg[\frac{I\{D=d\} \cdot (1 - p_{d0}(M,X))}{ {p_{d0}(M,X) \cdot (1 - p_{d0}(X))}}\cdot [Y- \mu_0(d,M,X)] \Bigg|X\Bigg] }^{=E[E[Y-\mu_0(d,M,X)| D=d, M, X] |D=1-d,X] = 0} \Bigg] \notag\\
&& + \ E\Bigg[\overbrace{  E\Bigg[  \frac{I\{D=1-d\}}{1- p_{d0}(X)}\cdot [ \mu_0(d,M,X) -   \omega_0(1-d,X)]   \Bigg| X \Bigg]}^{  =   E[\mu_0(d,M,X) -  \omega_0(1-d,X)|D=1-d,X] = 0 } \Bigg] \notag\\
&& + \  E[ \omega_0(1-d,X)] \ \ - \ \ \Psi_{d0}  \notag\\
& = &\Psi_{d0}\ \ - \ \ \Psi_{d0} \ \  =  0.  \notag
\end{eqnarray}
To better see this result, note that
\begin{eqnarray}
&&E\Bigg[ \frac{I\{D=d\} \cdot (1-p_{d0}(M,X)) }{ p_{d0}(M,X) \cdot (1-p_{d0}(X))}\cdot [Y-\mu_0(d,M,X)] \Bigg|X\Bigg]\notag\\
&=&E\Bigg[ E\Bigg[\frac{I\{D=d\}  }{ p_{d0}(M,X)  }\cdot [Y-\mu_0(d,M,X)]\Bigg|M,X\Bigg] \cdot  \frac{(1-p_{d0}(M,X))}{(1-p_{d0}(X))} \Bigg|X\Bigg]\notag\\
&=&E \Bigg[ E [Y-\mu_0(d,M,X) |D=d,M,X ] \cdot  \frac{(1-p_{d0}(M,X))}{(1-p_{d0}(X))}  \Bigg|X \Bigg]\notag\\
&=&E [ E [Y-\mu_0(d,M,X) |D=d,M,X ]   |D=1-d, X ]\notag\\
&=&E [ \mu_0(d,M,X)-\mu_0(d,M,X)   |D=1-d, X ]=0,\notag
\end{eqnarray}
where the first equality follows from the law of iterated expectations, the second from basic probability theory, and the third from Bayes' Law. Furthermore, 
\begin{eqnarray}
&&E\Bigg[  \frac{I\{D=1-d\}}{1-p_{d0}(X)}\cdot [ \mu_0(d,M,X) -  \omega_0(1-d,X)]   \Bigg| X \Bigg]\notag\\
&=&E\Bigg[ E\Bigg[\frac{I\{D=1-d\}}{1-p_{d0}(X)}\cdot [ \mu_0(d,M,X) -  \omega_0(1-d,X)] \Big| M, X \Bigg] \Bigg| X \Bigg]\notag\\
&=&E\Bigg[    [ \mu_0(d,M,X) -  \omega_0(1-d,X)] \cdot \frac{1-p_{d0}(M,X)}{1-p_{d0}(X)}  \Bigg| X \Bigg]\notag\\
&=&E [     \mu_0(d,M,X) -  \omega_0(1-d,X)   | D=1-d, X ]=E [     \mu_0(d,M,X) | D=1-d, X ] -  \omega_0(1-d,X)  \notag\\
&=& \omega_0(1-d,X)  -  \omega_0(1-d,X) =0, \notag
\end{eqnarray}
where the first equality follows from the law of iterated expectations and the third from Bayes' Law.

\textbf{Assumption 3.1(b)}

\textbf{Linearity:} The score $ \psi_d^*(W, \eta^*_0, \Psi_{d0}) $ is linear in $ \Psi_{d0}$ as it can be written as:
$\psi_d^*(W, \eta^*_0, \Psi_{d0}) = \psi_d^a(W, \Psi_{d0}) \cdot \Psi_{d0} + \psi_d^b(W, \eta^*_0) $
with $\psi_d^a(W, \eta^*_0) = -1$ and
\small
\begin{eqnarray}
\psi_d^b(W, \eta^*_0) &=& \frac{I\{D=d\} \cdot (1-p_{d0}(M,X))}{p_{d0}(M,X) \cdot (1-p_{d0}(X))} \cdot \Big[Y-\mu_0(d,M,X)\Big] \notag
\\ &+& \frac{I\{D=1-d\}}{1-p_{d0}(X)} \cdot \Big[\mu_0(d,M,X) - \omega_0(1-d,X) \Big] +  \omega_0(1-d,X) \notag
\end{eqnarray}
\\

\textbf{Assumption 3.1(c)}

\textbf{Continuity:}
The expression for the second Gateaux derivative of a map $\eta^* \mapsto E\Big[\psi^*_d(W, \eta^*, \Psi_{d0})\Big]$ is continuous.
\vspace{5pt}

\textbf{Assumption 3.1(d)}

\textbf{Neyman orthogonality}: \\

The Gateaux derivative in the direction\\ $ \eta^* - \eta^*_0 =
(\mu_{d}(d,M,X) - \mu_0(d,M,X), \omega(1-d,X) - \omega_0(1-d,X), p_{d}(M.X) - p_{d0}(M,X), p_{d}(X) - p_{d0}(X) ) $ is given by:
{\footnotesize
\begin{align}
	\partial E & \big[\psi_d^*(W, \eta^*,\Psi_{d})\big] \big[\eta^* - \eta^*_0 \big]
	 \notag \\
	 = & E\Bigg[ \frac{-[p_d(M,X) - p_{d0}(M,X)]}{p_{d0}(M,X)^2} \cdot \overbrace{\frac{I\{D = d\}}{1-p_{d0}(X)} \cdot \big(Y-\mu(d,M,X)\big)}^{E[ \cdot | X] =E [ Y-\mu(d,M,X)  | D=d,X  ]\cdot \frac{p_{d0}(X)}{1-p_{d0}(X)} = 0}\Bigg] \notag\\
	  +& E\Bigg[\overbrace{\frac{I\{D = d\} \cdot (1-p_{d0}(M,X))}{p_{d0}(M,X)\cdot (1-p_{d0}(X))} \cdot \big(Y-\mu_0(d,M,X)\big)}^{E[ \cdot | X]=E[E[Y-\mu_0(d,M,X)| D=d, M, X] |D=1-d,X] = 0} \cdot \frac{p_d(X)-p_{d0}(X)}{(1-p_{d0}(X))}\Bigg] \notag \\
	  + &  E\Bigg[\underbrace{\frac{I\{D = 1-d\}}{(1-p_{d0}(X))} \cdot   \big(\mu_0(d,M,X) - \omega_0(1-d,X) \big) }_{E[\cdot|X]=    E[\mu_0(d,M,X) -  \omega_0(1-d,X)|D=1-d,X] = 0 }  \cdot \frac{p_d(X) - p_{d0}(X)}{(1-p_{d0}(X))} \Bigg] \notag \\
	 &\underbrace{- E\Bigg[ \underbrace{\frac{I\{D = d\}}{p_{d0}(M,X)}}_{E[\cdot|M,X]=1} \cdot\frac{(1-p_{d0}(M,X))}{ (1-p_{d0}(X))} \cdot \Big[ \mu(d,M,X) - \mu_0(d,M,X)\Big]\Bigg] + E\Bigg[\underbrace{\frac{I\{D=1-d\}}{1-p_{d0}(X)}}_{E[\cdot|M,X]=\frac{1-p_{d0}(M,X)}{1-p_{d0}(X)}} \cdot \Big[\mu(d,M,X) -\mu_0(d,M,X)\Big]\Bigg]}_{= 0} \notag\\
	& \underbrace{- E\Bigg[\underbrace{\frac{I\{D=1-d\}}{1-p_{d0}(X)}}_{E[ \cdot | X] = 1}  \cdot \Big[\omega(1-d,X) -\omega_0(1-d,X)\Big] + \Big[\omega(1-d,X) -\omega_0(1-d,X)\Big]\Bigg]}_{= 0}. \notag
\end{align}
}
\raggedright
Thus, it follows that:
\begin{align}
&\partial E \big[\psi_d^*(W, \eta^*, \Psi_{d0})\big] \big[\eta^* - \eta^*_0 \big] = 0 \notag
\end{align}
proving that the score function is orthogonal.

\textbf{Assumption 3.1(e)}

\textbf{Singular values of $E[\psi^{a}_{d}(W;\eta^*_0)]$ are bounded:}
This holds trivially, because $\psi^{a}_{d}(W;\eta^*_0) = -1.$
\bigskip

\textbf{Assumption 3.2:  Score regularity and quality of nuisance parameter estimators}
\vspace{5pt}

Bounds for $m_n, m'_n, r_n, r'_n$ are omitted for the sake of brevity, because their derivations follow similarly as in the proof for $Y(d,M(1-d))$ in subsection \ref{Neyman}. However, the proof differs in establishing the bound on $\lambda'_n$ in 3.2(c) of \cite{Chetal2018}, as it is based on the regularity conditions in Assumption 5 that include $p_{d}(M,X)$ and $\omega(1-d,X)$.

\textbf{Bound for $\lambda'_n$:}
Consider
\begin{equation}
f(r) := E[\psi(W,\eta^*_0 + r(\eta^*-\eta^*_0),\Psi_{d0})] \notag
\end{equation}

We subsequently omit arguments for the sake of brevity and use $ \mu = \mu(d,M,X),  \omega = \omega(1-d,X), p_{d}= p_{d}(X),p_{dm}= p_{d}(M,X)$ and similarly $\mu_0, \omega_0,p_{d0},p_{dm0}.$


 For any $r \in (0,1):$
\begin{eqnarray} \label{secondGatDer}
\frac{\partial^2 f(r)}{\partial r^2}&=& E\Bigg[  (-2) \cdot I\{D = d\} \frac{(p_{dm} -p_{dm0})(p_{d} -p_{d0}) \left(Y-\mu_0 - r(\mu-\mu_0) \right)}{\left(p_{dm0} + r(p_{dm} -p_{dm0})\right) \left(1-p_{d0} + r(p_{d0} -p_{d})\right)^2}  \Bigg] \notag  \\
&+& E\Bigg[  2 \cdot I\{D = d\} \frac{(p_{dm} -p_{dm0})^2 \left(Y-\mu_0 - r(\mu-\mu_0) \right)}{\left(p_{dm0} + r(p_{dm} -p_{dm0})\right)^2 \left(1-p_{d0} + r(p_{d0} -p_{d})\right)}  \Bigg] \notag \\  
&+& E\Bigg[  2 \cdot I\{D = d\} \frac{\left(1-p_{dm0} + r(p_{dm0} -p_{dm})\right)(p_{d} -p_{d0}) (\mu-\mu_0) }{\left(p_{dm0} + r(p_{dm} -p_{dm0})\right) \left(1-p_{d0} + r(p_{d0} -p_{d})\right)^2}  \Bigg] \notag  \\ 
&+& E\Bigg[  (-2) \cdot I\{D = d\} \frac{ \left(1-p_{dm0} + r(p_{dm0} -p_{dm})\right)(p_{dm} -p_{dm0}) \left(Y-\mu_0 - r(\mu-\mu_0) \right)}{\left(p_{dm0} + r(p_{dm} -p_{dm0})\right)^2 \left(1-p_{d0} + r(p_{d0} -p_{d})\right)}  \Bigg] \notag \\   
&+& E\Bigg[  (-2) \cdot I\{D = d\} \frac{(p_{d} -p_{d0})^2 \left(1-p_{dm0} + r(p_{dm0} -p_{dm})\right) \left(Y-\mu_0 - r(\mu-\mu_0) \right)}{\left(p_{dm0} + r(p_{dm} -p_{dm0})\right) \left(1-p_{d0} + r(p_{d0} -p_{d})\right)^3}  \Bigg] \notag \\  
&+& E\Bigg[  (-2) \cdot I\{D = d\} \frac{(p_{dm} -p_{dm0}) (p_{d} -p_{d0}) \left(1-p_{dm0} + r(p_{dm0} -p_{dm})\right) \left(Y-\mu_0 - r(\mu-\mu_0) \right)}{\left(p_{dm0} + r(p_{dm} -p_{dm0})\right)^2 \left(1-p_{d0} + r(p_{d0} -p_{d})\right)^2}  \Bigg] \notag  \\ 
&+& E\Bigg[  (-2) \cdot I\{D = d\} \frac{(p_{dm} -p_{dm0})^2 \left(1-p_{dm0} + r(p_{dm0} -p_{dm})\right) \left(Y-\mu_0 - r(\mu-\mu_0) \right)}{\left(p_{dm0} + r(p_{dm} -p_{dm0})\right)^3 \left(1-p_{d0} + r(p_{d0} -p_{d})\right)}  \Bigg] \notag \\  
&+& E\Bigg[  2 \cdot I\{D = d\} \frac{(p_{dm} -p_{dm0})  (\mu-\mu_0)}{\left(p_{dm0} + r(p_{dm} -p_{dm0})\right) \left(1-p_{d0} + r(p_{d0} -p_{d})\right)}  \Bigg] \notag \\ 
&+& E\Bigg[  2 \cdot I\{D = 1-d\} \frac{(p_{d} -p_{d0})^2 \left( \mu_0  -  \omega_0 \right)}{\left(1-p_{d0} + r(p_{d0} -p_{d})\right)^3}  \Bigg] + E\Bigg[  2 \cdot I\{D = 1-d\} \frac{(p_{d} -p_{d0})^2 r \left(  (\mu-\mu_0)  - (\omega-\omega_0) \right)}{\left(1-p_{d0} + r(p_{d0} -p_{d})\right)^3}  \Bigg] \notag \\
&+& E\Bigg[  2 \cdot I\{D = 1-d\} \frac{(p_{d} -p_{d0}) \left( \mu-\mu_0 \right)}{\left(1-p_{d0} + r(p_{d0} -p_{d})\right)^2}  \Bigg] + E\Bigg[  (-2) \cdot I\{D = 1-d\} \frac{(p_{d} -p_{d0}) \left( \omega-\omega_0 \right)}{\left(1-p_{d0} + r(p_{d0} -p_{d})\right)^2}  \Bigg]  \notag  
\end{eqnarray}

Bounding these twelve terms proceeds similarly as in subsection \ref{Neyman}. In order to bound the eighth term, we make use of the sixth inequality in \ref{Tnstar}. Similarly, for bounding the tenth and the twelfth terms we make use of the last inequality in \ref{Tnstar}. Thus, we get that for some $C_{\epsilon}''$ that only depends on $C$ and $\epsilon$
\begin{equation}
\left|\frac{\partial^2 f(r)}{\partial r^2} \right| \leq C_{\epsilon}'' \delta_n n^{-1/2} \leq \delta_n' n^{-1/2}. \notag
\end{equation}
This provides the upper bound on $\lambda'_n$ in Assumption 3.2(c) of \cite{Chetal2018} as long as $C_{\epsilon} \geq C_{\epsilon}''$.

This concludes the proof of Theorem 2. $\hfill\square$

\end{appendix}
\end{document}